%% file: main.tex
\documentclass{article} 
\usepackage{iclr2027_conference,times}

\input{math_commands.tex}

\usepackage{hyperref}
\usepackage{url}
\usepackage{booktabs}
\usepackage{multirow}
\usepackage{enumitem}
\usepackage{graphicx}
\usepackage{makecell}
\usepackage{xcolor}
\usepackage[table]{xcolor}
\newcommand{\metricpair}[2]{%
  \shortstack[c]{#1\\[-1pt]#2}%
}

\title{\includegraphics[height=0.8em]{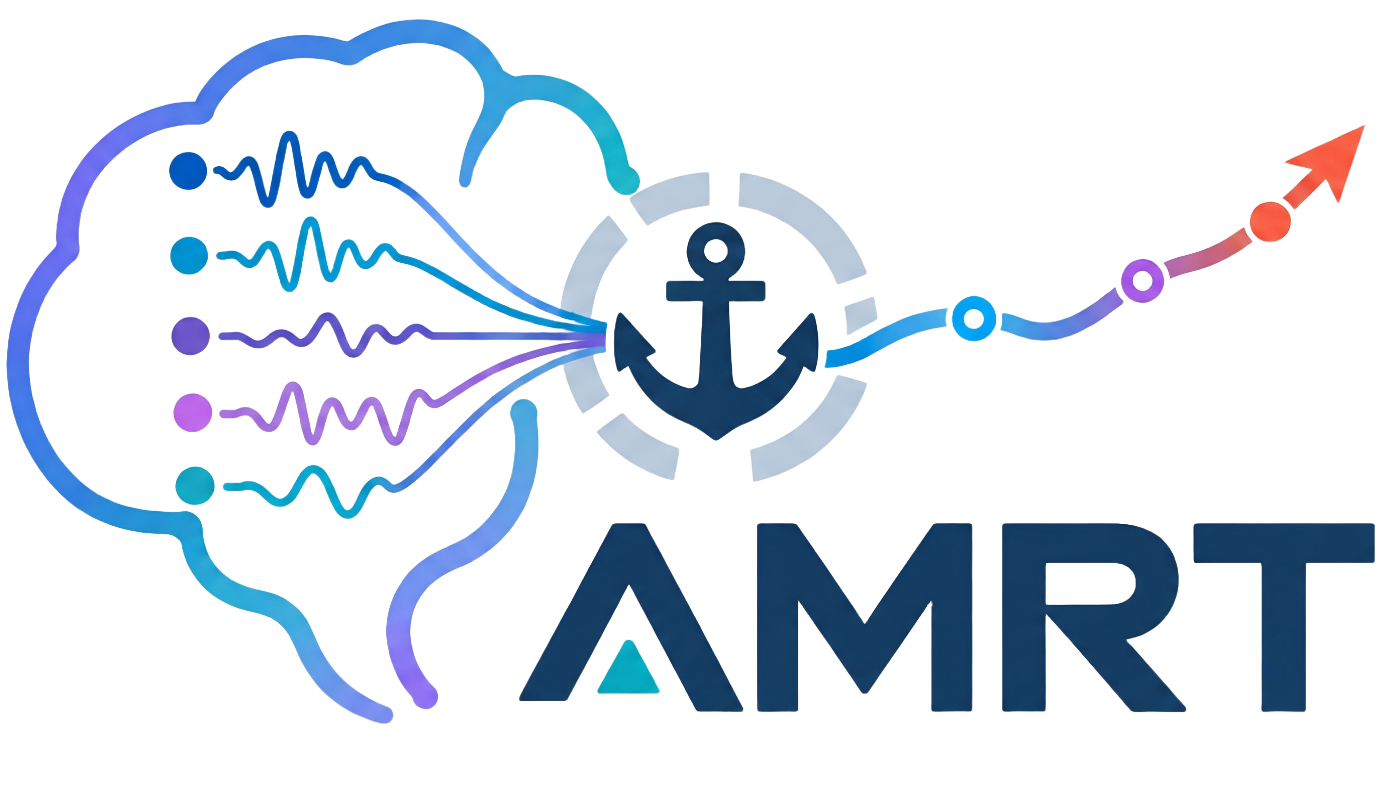} Anchoring Motion, Refining Trajectories: Generalizable Continuous Neural Decoding}

\author{
Fanqi Shen$^{1}$, Junru Chen$^{1}$, Surui Cheng$^{2}$,
Jiahe Li$^{1}$, ZhiZhang Yuan$^{1}$, Gaorui Zhang$^{1}$, \\
ZhiYuan Cao$^{1}$, Xiaoquan Shen$^{1}$, Yang Yang$^{1,*}$ \\[0.6em]
$^{1}$Zhejiang University, Hangzhou, Zhejiang, China \\
$^{2}$Columbia University, New York, NY, USA
}

\newcommand{\model}{AMRT}

\iclrfinalcopy 
\begin{document}

\maketitle

\begin{abstract}
\input{paper/abstract}
\end{abstract}

\input{paper/introduction}
\input{paper/methodology}

\input{paper/experiment}
\input{paper/conclusion}

\input{paper/statement}

\bibliography{iclr2027_conference}
\bibliographystyle{iclr2027_conference}

\clearpage
\appendix
\input{appendix/related}
\input{appendix/baseline}
\input{appendix/dataset}
\input{appendix/model}
\input{appendix/experiment}

\end{document}

%% file: math_commands.tex
\usepackage{amsmath,amsfonts,bm}

\def\eqref#1{equation~\ref{#1}}

\def\1{\bm{1}}

\DeclareMathAlphabet{\mathsfit}{\encodingdefault}{\sfdefault}{m}{sl}
\SetMathAlphabet{\mathsfit}{bold}{\encodingdefault}{\sfdefault}{bx}{n}



%% file: paper/abstract.tex
Brain activity encodes motor information across multiple temporal scales, enabling continuous neural control beyond discrete action recognition.
However, existing methods are typically tailored to specific subjects and tasks, with direct neural-to-kinematic mappings that can entangle subject-specific neural variability with task-dependent motion patterns, limiting generalization.
To address these limitations, we introduce \model{}, a shared neural decoding framework for heterogeneous electrode configurations and movement paradigms. Motivated by cortical encoding of movement direction and speed, \model{} adopts a two-stage design that anchors local motion through velocity estimation and refines the resulting trajectories over longer horizons, encouraging reuse of local motion tendencies across subjects and tasks.
Across three datasets, \model{} generalizes to unseen trials and subjects and demonstrates transfer potential across neural recordings and movement paradigms, highlighting the potential of shared neural representations and anchor-and-refine decoding for generalizable continuous neural control. 

%% file: paper/introduction.tex
\section{Introduction}
Brain-computer interfaces (BCIs) translate neural activity into control commands for external devices, with motor decoding forming the basis of applications \citep{xu2021review}. Electrophysiological motor signals can be recorded from scalp electroencephalography (EEG) or intracranial EEG (iEEG), providing complementary access to movement-related neural activity. 
Most existing motor-decoding studies focus on \textbf{discrete decoding}, classifying imagined or executed movements into predefined action categories \citep{srinivasan2025motor, 
ju2024deep, shen2026brain4fms}.
While these paradigms establish the decodability of motor-related EEG patterns, discrete decisions alone cannot support the moment-to-moment adjustments required for real-world motor control. Continuous decoding instead recovers time-varying motor commands, making it essential for fine-grained control of cursors, robotic effectors, and neuroprostheses \citep{pandarinath2022science}. 

\citet{korik2018decoding} showed that EEG signals contain sufficient information to reconstruct continuous hand trajectories, establishing the feasibility of continuous motor decoding. 
Building on this foundation, recent advances in deep temporal modeling, including recurrent networks, Transformers, and multi-scale attention mechanisms, have substantially improved continuous kinematic decoding, enabling increasingly accurate reconstruction of hand velocity and trajectories from electrophysiological brain signals \citep{duan2025eeg, giri2026neurokinect, li2026copilot, zhu2026m3t}. 
Yet robust continuous decoding across subjects and tasks remains challenging. \textbf{(1) Neural distribution shift:} neural signals vary substantially across subjects, sessions, and recording systems, while heterogeneous electrode configurations further hinder representation alignment. \textbf{(2) Decoding-target design:} existing methods typically regress position or velocity directly \citep{kwon2026e2t, mc2023decoding, zhu2026m3t}, making the decoding target sensitive to differences in movement workspace and task structure. Position preserves global trajectory structure but is closely tied to task-specific workspaces and initial states, whereas velocity describes local movement dynamics but accumulates integration error over time.  \textbf{(3) Generalization scope:} existing studies largely evaluate within fixed subjects, recording configurations, or datasets, leaving generalization across these settings insufficiently understood. 
These challenges motivate a unified framework for robust and transferable continuous neural decoding.

These considerations are consistent with the temporal organization of motor-cortical activity.
Motor cortex encodes movement direction and speed \citep{moran1999motor}, making velocity a natural representation of local motion. Meanwhile, cortical population dynamics unfold over longer timescales \citep{churchland2010cortical,kaufman2014cortical}.
This motivates separating local motion estimation from longer-horizon trajectory reconstruction rather than learning a single direct neural-to-position mapping.
We therefore introduce \textbf{\model} (\textbf{A}nchoring \textbf{M}otion, \textbf{R}efining \textbf{T}rajectories), a two-stage framework that uses velocity as an intermediate motion representation. 
Stage-0 predicts local motion tendency from neural context and integrates it into a coarse trajectory, providing a motion anchor that captures the local movement direction and approximate progression. Stage-1 then combines the coarse trajectory with temporally extended neural context to predict residual trajectory corrections, explicitly compensating for integration drift and recovering long-horizon trajectory structure.
To handle neural heterogeneity across subjects and recording configurations, AMRT further maps different electrode layouts into a shared representation through a channel adapter. Together, channel adaptation and anchor-and-refine decoding address heterogeneity in the neural and motion spaces, respectively.
We evaluate \model{} from unseen trials and subjects to cross-dataset transfer across heterogeneous EEG/iEEG recordings. Across three datasets, AMRT generalizes under increasingly challenging shifts, while transfer analyses reveal that reusable information can emerge at different levels of the decoding hierarchy.
Our contributions are three-fold:
\begin{enumerate}[leftmargin=*]
\item \textbf{An anchor-and-refine formulation for generalizable continuous decoding.}
We separate local motion tendency estimation from long-horizon trajectory reconstruction, using velocity to anchor moment-to-moment motion dynamics and trajectory-space residual refinement to correct accumulated drift. This decomposition provides a structured intermediate representation for transferring local motion while preserving trajectory-level consistency.
\item \textbf{A shared neural decoding framework for heterogeneous domains.}
We introduce AMRT, which combines a shared neural encoder with lightweight channel adaptation and a common action-chunk decoder, enabling continuous decoding across subjects and datasets with heterogeneous electrode configurations and movement paradigms.
\item \textbf{A hierarchical evaluation of continuous-decoding generalization.} We evaluate generalization from unseen trials and subjects to transfer across datasets with different recording configurations and movement paradigms. To our knowledge, this is the first systematic study of cross-dataset transfer for continuous kinematic decoding across heterogeneous EEG/iEEG recording settings.
\end{enumerate}

%% file: paper/methodology.tex
\section{Methodology}

\begin{figure}[t]
    \centering
    \includegraphics[width=0.95\linewidth]{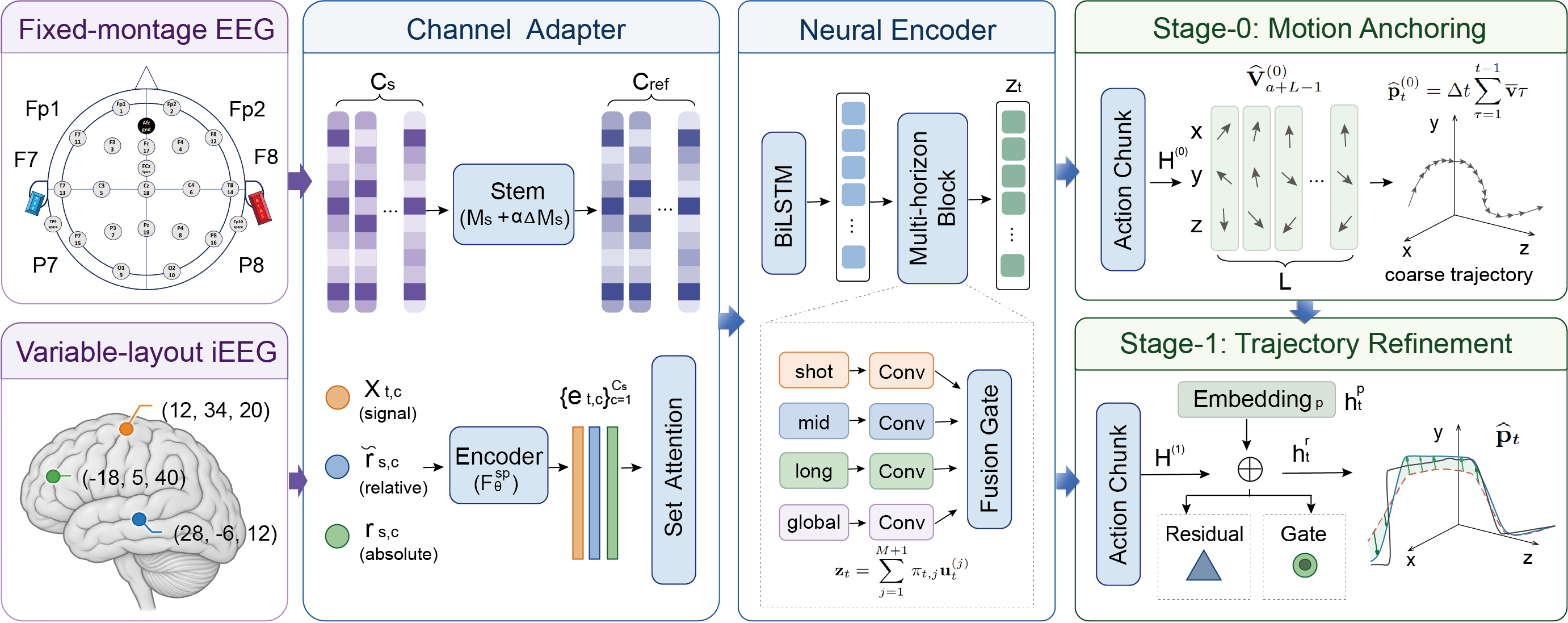}
    \caption{Overview of the Anchoring Motion, Refining Trajectories (\model{}) framework. 
    }
    \label{fig:main}
\end{figure}

\label{sec:method}
This section presents {\model} for continuous kinematic decoding from heterogeneous neural recordings. After formulating the problem, we describe heterogeneous neural alignment for shared decoding in Section~\ref{sec:alignment}, followed by motion anchoring and trajectory refinement in Section~\ref{sec:amrt}.

\subsection{Problem Formulation}
\label{sec:form}
We consider continuous kinematic decoding from complete movement trials. Let the $n$-th complete trial of subject $s$ be
$\mathcal{D}_{s,n}
    = \left\{(\mathbf{X}_t,\mathbf{p}_t,\mathbf{v}_t)\right\}_{t=1}^{T_{s,n}}$,
where $T_{s,n}$ is the number of control steps. At step $t$, $\mathbf{X}_t\in\mathbb{R}^{C_s\times P}$ is the neural segment aligned with the corresponding control step, where $C_s$ is the subject-specific number of electrodes and $P$ is the number of neural samples per step. The position and velocity are $\mathbf{p}_t,\mathbf{v}_t\in\mathbb{R}^{D}$, where $D\in\{2,3\}$ depends on the movement space of the dataset. Velocity labels are derived from the recorded position trajectories during preprocessing. To reduce trial-specific dependence on the initial hand position, we decode trajectories in relative coordinates, using $\mathbf{p}^{\mathrm{gt}}_t=\mathbf{p}_t-\mathbf{p}_1$ as the trajectory-space target.

To provide sufficient temporal context while retaining online compatibility, we perform decoding over consecutive output blocks rather than individual control steps. For a block of length $L$ starting at step $a$, the model observes an $H$-step neural context ending at the block endpoint:

\vspace{-1.5em}
{
\footnotesize
\begin{equation}
    \mathcal{X}_{a}^{(H)}
    =
    \mathbf{X}_{a+L-H:a+L-1},
    \qquad H \geq L,
    \label{eq}
\end{equation}
}
\vspace{-2em}

where successive blocks advance by a stride $S$. Boundary contexts are edge-padded and excluded from the training objective using a validity mask. This formulation is block-causal: no neural activity after step $a+L-1$ is used, while observations within the current block are jointly available for prediction. The same temporal protocol is used throughout the framework.

\subsection{Heterogeneous Neural Alignment for Shared Decoding}
\label{sec:alignment}
This subsection introduces a shared decoding architecture for heterogeneous neural recordings. It combines modality-aware channel adaptation, multi-horizon neural encoding, and query-based action-chunk decoding to map EEG/iEEG signals into motion predictions.
\subsubsection{Modality-aware channel adapter}
Neural recordings exhibit distinct forms of spatial heterogeneity. EEG typically follows standardized montages with stable channel identities, whereas iEEG uses subject-specific electrode sets without consistent index correspondence across patients. We therefore adopt two complementary channel adapters: an \textbf{exact channel adapter} for fixed-montage recordings and a \textbf{geometry-aware channel adapter} for variable electrode layouts. In modality-specific experiments, we use the corresponding adapter for EEG and iEEG; for cross-dataset transfer, the geometry-aware adapter provides a shared interface that decouples downstream decoding from dataset-specific electrode indexing.

\paragraph{Exact projection for fixed montages.}
Standardized EEG montages provide reproducible electrode layouts and stable channel identities, enabling reliable channel correspondence across subjects \citep{hh1958ten, tucker1993spatial}. We canonicalize the labels and construct a deterministic matrix $\mathbf{M}_{s}\in\mathbb{R}^{C_{\mathrm{ref}}\times C_s}$ for montage, where $\mathcal{C}_s$ is the number of valid channels. Each observed channel is mapped to its matching reference-montage entry, with unmatched entries set to zero. We optionally augment this nominal correspondence with a small learnable residual:

\vspace{-1.4em}
{
\footnotesize
\begin{equation}
    \widetilde{\mathbf{X}}_t
    =\left(\mathbf{M}_s+\alpha\,\Delta\mathbf{M}_s\right)\mathbf{X}_t,
    \label{eq:exact_adapter}
\end{equation}
}
\vspace{-1.6em}

where $\Delta\mathbf{M}_s$ is initialized to zero and $\alpha$ controls its scale. 
The residual is shared across source subjects to correct errors beyond the fixed canonical mapping. It is frozen for zero-shot transfer and adapted only with labeled target data, enabling lightweight cross-subject adaptation.

\paragraph{Geometry-aware encoding for variable electrode sets.}
Unlike EEG, iEEG varies across subjects in channel count and cortical coverage \citep{holdgraf2019ieeg}, making channel correspondence unreliable. We therefore treat valid contacts as a coordinate annotated set, where a shared encoder embeds each electrode segment and coordinate encoders incorporate centered local geometry and global implantation location. 
Let $\mathbf{r}_{s,c}\in\mathbb{R}^{3}$ denote the coordinate of electrode $c$ for subject $s$. To obtain a subject-normalized representation of the local electrode layout $\widetilde{\mathbf{r}}_{s,c}$, we center the coordinates and normalize their scale by the RMS spatial radius \citep{gower1975generalized, qi2017pointnet}. 
Specifically, for electrode coordinate $\mathbf{r}_{s,c}$, we use $\widetilde{\mathbf{r}}_{s,c}=(\mathbf{r}_{s,c}-\boldsymbol{\mu}_s)/(\rho_s+\epsilon)$, where $\boldsymbol{\mu}_s$ and $\rho_s$ denote the mean coordinate and RMS radius of the valid electrodes, respectively.
For each electrode, we jointly encode its neural activity, subject-normalized relative geometry, and absolute spatial location:

\vspace{-1.4em}
{
\small
\begin{equation}
    \mathbf{e}_{t,c}
    =
    F_{\theta}^{\mathrm{sp}}
    \left(
        \mathbf{X}_{t,c},
        \widetilde{\mathbf{r}}_{s,c},
        \mathbf{r}_{s,c}
    \right).
\end{equation}
}
\vspace{-1.6em}

The normalized coordinates capture the relative electrode configuration, while the original coordinates retain the absolute recording location removed by canonicalization. 
These geometry-aware electrode embeddings form an unordered set. We aggregate them using a small set of learned spatial queries with masked multi-head attention~\citep{lee2019set}, followed by a fusion layer to obtain the shared neural representation $\mathbf{h}_t$. Invalid or padded channels are excluded by the attention mask. This set-based aggregation is permutation invariant and naturally accommodates different electrode counts and layouts without assuming cross-subject channel identity.


\subsubsection{Multi-horizon neural encoder}
Movement-related neural activity spans multiple temporal scales. We therefore encode the channel-aligned context using complementary local and global temporal operators. After an initial temporal stem and a bidirectional recurrent layer that captures sustained trends, the sequence is processed by stacked multi-horizon blocks. 
Given the same input sequence, each block applies $J$ temporal convolution branches with different receptive fields and one global attention branch, producing branch features $\mathbf{u}_t^{(j)}\in\mathbb{R}^{d}$ at control step $t$. We adaptively fuse these features with a learned gating network:

\vspace{-1.6em}
{
\small
\begin{equation}
\boldsymbol{\pi}_t
=
\operatorname{softmax}
\left(
\phi_{\pi}
\left(
[\mathbf{u}^{(1)}_t;\ldots;\mathbf{u}^{(J+1)}_t]
\right)
\right),
\qquad
\mathbf{z}_t
=
\sum_{j=1}^{J+1}
\pi_{t,j}\mathbf{u}^{(j)}_t,
\label{eq}
\end{equation}
}
\vspace{-1.5em}

where $\phi_{\pi}$ outputs $J+1$ fusion weights, $\mathbf{u}^{(1)}_t,\ldots,\mathbf{u}^{(J)}_t$ are features from the temporal branches, and $\mathbf{u}^{(J+1)}_t$ is the global-attention feature. Since the fusion weights are conditioned on the current branch responses, the model can emphasize different temporal scales at different control steps. Stacking the fused features over time yields the memory $\mathbf{Z}=[\mathbf{z}_1,\ldots,\mathbf{z}_H]\in\mathbb{R}^{H\times d}$, which combines local neural dynamics with longer-range movement context.
Multi-scale temporal modeling has proven effective for EEG decoding~\citep{altaheri2023dynamic, zhu2026m3t}, while convolution and attention provide complementary local and global temporal biases~\citep{song2022eeg}.
All operations are restricted to the $H$-step context defined in Sec. \ref{sec:form}, preserving the block-causal protocol.

\subsubsection{Query-based action-chunk decoder}
Rather than decoding each control step independently, we jointly predict an $L$-step motion chunk. Following the action-chunking paradigm \citep{zhao2023learning}, each relative position is assigned a learned query. The queries interact through self-attention and attend to the encoded neural memory, enabling joint prediction across the chunk. For stage $s\in\{0,1\}$, we define stage-specific temporal queries $\mathbf{Q}^{(s)}=[\mathbf{q}^{(s)}_1,\ldots,\mathbf{q}^{(s)}_L]$ and compute 
$\mathbf{H}^{(s)}=\operatorname{ChunkDec}^{(s)}\left(\mathbf{Q}^{(s)},\mathbf{Z}\right)$, where $\mathbf{H}^{(s)}=[\mathbf{h}^{(s)}_1,\ldots,\mathbf{h}^{(s)}_L]$ is the chunk representation used by the corresponding stage. $\operatorname{ChunkDec}^{(s)}$ uses self-attention over queries and cross-attention to $\mathbf{Z}$. Each query corresponds to a fixed temporal position in the output chunk, so no matching is required \citep{carion2020end}. 
Each $\operatorname{ChunkDec}^{(s)}$ uses a single sequence of query self-attention, cross-attention to $\mathbf Z$, a residual connection from the original learned queries followed by LayerNorm, and a position-wise MLP. Appendix~D.1 gives the sublayer equations.

\subsection{Motion Anchoring and Trajectory Refinement}
\label{sec:amrt}
Given the decoded chunk representation, we reconstruct continuous trajectories in two stages. Stage-0 represents local motion with velocity and integrates it into a coarse trajectory, while Stage-1 refines long-horizon trajectory structure through residual correction.

\subsubsection{Stage-0: Motion Anchoring}
Stage-0 captures local movement dynamics by predicting an $L$-step velocity chunk from the decoded representation: $\widehat{\mathbf{V}}^{(0)}_{a:a+L-1}=\phi_{0}\!\left(\mathbf{H}^{(0)}\right)$, where $\phi_{0}$ is the velocity head. We refer to $\widehat{\mathbf{V}}^{(0)}$ as \textbf{motion anchors}, as they provide locally coherent estimates of movement direction and speed for subsequent trajectory reconstruction. 
Stage-0 is optimized with a masked Huber loss \citep{huber1992robust} over valid velocity targets, with dynamic per-axis reweighting to balance decoding difficulty across motion axes.
After training, Stage-0 remains frozen during trajectory refinement.

As movement unfolds, Stage-0 produces overlapping velocity anchors that are integrated into a coarse trajectory. Predictions are first denormalized using training-set statistics, then combined by Hann-weighted overlap-add and integrated, yielding the velocity estimate $\overline{\mathbf{v}}_t$. The coarse trajectory is then obtained by cumulatively integrating these estimates from the beginning of the trial:

\vspace{-1.6em}
{
\footnotesize
\begin{equation}
\widehat{\mathbf{p}}^{(0)}_1=\mathbf{0},
\qquad
\widehat{\mathbf{p}}^{(0)}_t
=
\Delta t
\sum_{\tau=1}^{t-1}
\overline{\mathbf{v}}{\tau},
\quad t>1,
\label{eq}
\end{equation}
}
\vspace{-1.5em}

where $\Delta t=1/f_c$ and $f_c$ is the control sampling rate. Thus, $\widehat{\mathbf{p}}^{(0)}_t$ summarizes the motion anchors accumulated up to the current step and provides the trajectory state that Stage-1 subsequently refines. Errors in the local velocity estimates naturally accumulate through this integration, making long-horizon drift explicit in trajectory space.

\subsubsection{Stage-1: Trajectory Residual Refinement}
Stage-0 reconstructs a coarse trajectory from local motion anchors, but velocity errors accumulate into long-horizon drift. Stage-1 therefore shifts the decoding objective from local motion to trajectory-space correction.
Rather than predicting trajectories from scratch,  Stage-1 learns how the coarse trajectory should be corrected. A separate refinement decoder combines the frozen neural memory with the current coarse trajectory state. 
For each control step, the refinement module conditions on both the coarse trajectory state and the corresponding neural representation. Specifically, we first embed the coarse position and fuse it with the refinement feature:

\vspace{-1.5em}
{
\small
\begin{equation}
\mathbf{h}^{r}_{t}
=
\operatorname{Fuse}
\left(
\mathbf{h}^{(1)}_{t},
\mathbf{h}^{p}_{t}
\right),
\qquad
\label{fuse}
\mathbf{h}^{p}_{t} = \operatorname{Emb}_{p}
\left(
\widehat{\mathbf{p}}^{(0)}_{t}
\right)
\end{equation}
}
\vspace{-1.6em}

where $\widehat{\mathbf{p}}^{(0)}_{t}$ is the coarse position, $\operatorname{Emb}_{p}$ maps the position into the feature space, and $\mathbf{h}^{(1)}_{t}$ provides the neural evidence used for trajectory refinement.
From the fused representation, we predict both a candidate residual correction and a gate that modulates how strongly this correction is applied:

\vspace{-1.3em}
{
\small
\begin{equation}
    \Delta\mathbf{r}_{t}
    =
    f_{\Delta}\left(\mathbf{h}^{r}_{t}\right),
    \qquad
    \mathbf{g}_{t}
    =
    \sigma\left(
    f_{g}\left(\mathbf{h}^{r}_{t}\right)
    \right).
\end{equation}
}
\vspace{-1.7em}

Here, $f_{\Delta}$ predicts the direction and magnitude of the trajectory correction, while $f_g$ determines how strongly the correction is applied at the current control step. We refer to the resulting residual as a \textbf{gated correction increment}. Since adjacent decoding blocks overlap in time, predictions associated with the same control step are first aggregated by overlap-add:

\vspace{-1.4em}
{
\small
\begin{equation}
\overline{\Delta\mathbf{r}}_{t}
=
\operatorname{OA}
\left(
\mathbf{g}_{t}\odot\Delta\mathbf{r}_{t}
\right),
\qquad
\mathbf{r}_{t}
=
\sum_{\tau=1}^{t-1}
\overline{\Delta\mathbf{r}}_{\tau},
\qquad
\widehat{\mathbf{p}}_{t}
=
\widehat{\mathbf{p}}^{(0)}_{t}
+
\mathbf{r}_{t},
\label{eq:trajectory_refinement}
\end{equation}
}
\vspace{-1.4em}

where $\operatorname{OA}(\cdot)$ denotes overlap-add aggregation across blocks, and $\mathbf{r}_{t}$ is the accumulated trajectory correction up to control step $t$. Predicting correction increments rather than absolute positions preserves the coarse motion structure established by Stage-0, while allowing long-horizon neural evidence to progressively compensate for accumulated trajectory errors. The gate allows the model to modulate correction magnitude. 
Accordingly, Stage-1 is optimized at the trial level with trajectory reconstruction, gap, smoothness, and gating-utility losses.
This two-stage design directly realizes our \textbf{anchor-and-refine} principle: Stage-0 establishes transferable local motion anchors, while Stage-1 corrects accumulated drift and restores trajectory-level consistency.



%% file: paper/experiment.tex
\section{Experiment}
We organize our experiments into a hierarchy of increasingly challenging generalization settings, from unseen trials to unseen subjects and cross-dataset transfer. Section~\ref{sec:setup} describes the evaluation setup. Section~\ref{sec:generalization} evaluates within-dataset generalization to unseen trials and subjects, while Section~\ref{sec:cross_dataset} extends the evaluation across heterogeneous datasets and recording settings and movement paradigms. Section~\ref{sec:motion_to_trajectory} further analyzes the motion-to-trajectory design.

\subsection{Experimental Setup}
\label{sec:setup}
\paragraph{Datasets.} We evaluate our framework on three public continuous motor decoding datasets spanning different movement paradigms and neural recording modalities. WAY-EEG-GAL \citep{luciw2014multi} contains EEG and 3D kinematics during grasp-and-lift movements under varying object conditions. We use wrist trajectory as the decoding target. U-Limb \citep{averta2021u} provides synchronized EEG and motion capture during diverse upper-limb movements, from which we decode the hand-dorsum trajectory. Joystick-Track, included in the ECoG library of \citet{miller2019library} and originally introduced by \citet{schalk2007decoding}, contains intracranial recordings during 2D joystick tracking. Together, these datasets span diverse upper-limb movements across EEG and iEEG recordings.

\paragraph{Preprocessing.} We convert all datasets into temporally aligned neural–motor sequences. Neural signals are resampled to 500 Hz, notch-filtered at 50 or 60 Hz, common-average referenced, and channel-wise z-score normalized within each chunk. We follow the original trajectory construction and coordinate processing for each dataset. Velocity is causally estimated from position using a third-order trailing Savitzky-Golay derivative \citep{savitzky1964smoothing}, while the original trajectory is retained as the position target. For each dataset and kinematic axis, velocity is standardized using training-set statistics: $\widetilde{\mathbf v}_t=(\mathbf v_t-\boldsymbol{\mu}_d)\oslash
\boldsymbol{\sigma}_d$, where $\boldsymbol{\mu}_d$ and $\boldsymbol{\sigma}_d$ are the per-axis mean and standard deviation computed from valid training labels.

\paragraph{Baselines and implementation.}
We compare our method with three recent continuous EEG decoding approaches: M3T-Attention \citep{zhu2026m3t}, Copilot-Assisted \citep{li2026copilot}, and the Transformer-Based Regressor (TBR) \citep{dangi2026towards}. All methods were originally evaluated on WAY-EEG-GAL. We reproduce each baseline following its reported preprocessing pipeline, model architecture, and training configuration as closely as possible, while evaluating all methods using the same data partitions and evaluation metric as our framework. 
We report trial-level trajectory Pearson correlation coefficient (\texttt{traj\_PCC}), computed between predicted and ground-truth trajectories over time and averaged across test trials. Unless specified otherwise, results are reported as mean$\pm$std across runs. Statistical significance is assessed using a two-sided paired $t$-test against the best-performing baseline, with *, **, and *** denoting $p<0.05$, $p<0.01$, and $p<0.001$.

\subsection{Generalization Across Trials and Subjects}
\label{sec:generalization}
We first assess AMRT at two levels of generalization, from unseen trials of observed subjects to unseen subjects with potentially shifted neural distributions and electrode configurations. We begin with cross-trial decoding, then evaluate cross-subject generalization under leave-one-subject-out (LOSO) setting and further examine the benefit of few-shot calibration.

\subsubsection{Cross Trial Decoding}
We begin with the first level of our generalization hierarchy, evaluating unseen trials under the \texttt{ctri-mix} protocol, which pools trials across subjects to mitigate the instability of within-subject evaluation when per-subject trials are scarce. Complete trials are pooled across subjects and split into training, validation, and test sets at an 8:1:1 ratio, with splitting performed at the trial level to avoid window leakage.  Results are aggregated over 10 random seeds. For fixed-channel baselines, we introduce \texttt{JT-common}, truncating Joystick-Track recordings to a common channel count across subjects. We additionally report within-subject \texttt{ctri} results on WAY-EEG-GAL for comparison with prior work \citep{jain2023eeg,jain2025esi,jain2023subject,li2026copilot}, and on Joystick-Track to evaluate decoding with all available channels retained.

\input{tables/ctri-mix}

Across the cross-trial settings, AMRT consistently achieves the highest PCC, with its advantage becoming more pronounced as the decoding setting becomes more heterogeneous (Table~\ref{tab:unseen_trial}). On WAY-EEG-GAL, where strong baselines already reach high PCC, AMRT provides a modest but consistent improvement under both \texttt{ctri-mix} and \texttt{ctri}. The gap widens substantially on U-Limb, where diverse upper-limb movements increase behavioral variability, with AMRT improving PCC from 0.314 to 0.530 over the best baseline. This advantage also extends to Joystick-Track, indicating that AMRT remains effective under a distinct iEEG recording setting. We further examine the contribution of individual model components under the \texttt{ctri-mix} protocol in Appendix~\ref{app:model-abl}.

\subsubsection{Generalization to Unseen Subjects}
We then evaluate cross-subject generalization under a subject-disjoint setting, where the test subject is not observed during training and may exhibit substantially different neural signal distributions. We first assess AMRT under LOSO evaluation and then examine whether limited target-subject calibration can provide further gains.

\paragraph{Cross-subject Decoding under LOSO.}
We adopt a leave-one-subject-out protocol (\texttt{csub}). For each fold, one subject is held out for testing, a disjoint subject is used for validation and checkpoint selection, and all remaining subjects are used for training. The procedure is repeated until every subject has served as the test subject once. Results are macro-averaged across held-out subjects, with no test-subject data used for training or model selection.
As shown in Table~\ref{tab:unseen_sub}, AMRT achieves the highest PCC across all three datasets. On WAY-EEG-GAL, strong baselines remain competitive on average but vary considerably across held-out subjects, indicating less consistent transfer. AMRT achieves stronger and more stable performance across all datasets, including Joystick-Track, where subject-dependent electrode configurations prevent fixed-channel baselines from being directly applied. We further ablate the channel adapter. AMRT w/o CA still outperforms the baselines, but consistently underperforms the full model in cross-subject decoding. This suggests that the shared decoder captures much of the transferable motion structure, while channel alignment provides a complementary improvement by mitigating subject-specific differences in electrode configurations. 

\input{tables/csub}

\paragraph{Few-shot calibration.}
We next investigate whether limited target-subject supervision can further improve the LOSO model. For each held-out subject, we use $K\in\{1,3,5,10\}$ labeled trials for calibration, where one shot corresponds to one complete trial. Calibration sets are nested across shot counts, and all models are evaluated on the same blind test set. We compare full fine-tuning (\textbf{Full-FT}), which updates Stage-0 and subsequently Stage-1, with \textbf{Stage-1-only adaptation}, which freezes the channel adapter and Stage-0 and updates only the trajectory-refinement module. Both strategies use a fixed adaptation schedule without target-subject checkpoint selection. Their comparison indicates whether limited calibration mainly benefits motion estimation or trajectory refinement.

As shown in Figure~\ref{fig:exp}(a) and Table~\ref{tab:fewshot_full}, the effect depends on both the zero-shot baseline and calibration size. On WAY-EEG-GAL, where LOSO performance is already strong, calibration yields only marginal gains. On U-Limb, performance improves progressively with more target trials, with Full-FT becoming increasingly beneficial as supervision grows. On Joystick-Track, Stage-1-only adaptation consistently improves over the LOSO model, whereas Full-FT is less stable in the low-shot regime. Overall, Stage-1-only adaptation is more reliable when calibration data are scarce, suggesting that limited supervision is often better used for trajectory refinement than for relearning the neural-to-motion mapping. As more target data become available, updating Stage-0 becomes increasingly beneficial in settings with larger subject-specific shifts. 

\subsection{Transfer Across Heterogeneous Datasets}
\label{sec:cross_dataset}
We next investigate whether AMRT can reuse learned representations across datasets with different recording modalities, electrode configurations, and movement paradigms. We focus primarily on single-source transfer to expose source-target compatibility, while dual-source transfer and heterogeneous-data scaling analyses are reported in Appendix~\ref{app:cdata-app}. 

For each target dataset, we construct repeated subject-disjoint training/validation/test splits with an approximate $8{:}1{:}1$ ratio, adjusted to the available cohort size, and repeat the procedure over six random seeds. Within each split, we subsample $f\in\{0.10,0.25,0.50,1.00\}$ of the target training trials while keeping the validation and test sets fixed. Velocity denormalization uses control statistics from the full target training split in all settings. At each fraction, we compare training from scratch with transfer. The \textbf{scratch} model is trained only on the selected target subset, whereas the \textbf{transfer} model is first pretrained on the source dataset and then adapted to the same target subset. We use both Full-FT and Stage-1-only adaptation, following the cross-subject setting. U-Limb, WAY-EEG-GAL, and Joystick-Track are abbreviated as U, W, and J.

\begin{figure}[t]
    \centering
    \includegraphics[width=1.0\linewidth]{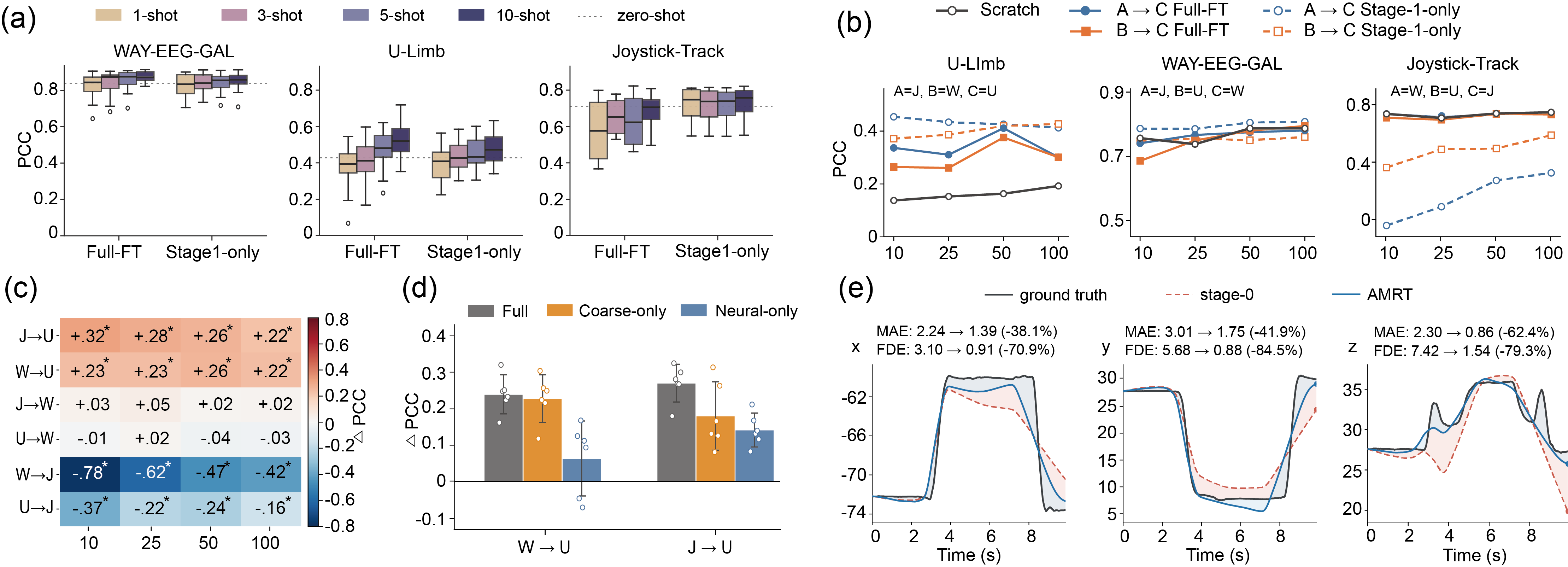}
    \caption{\textbf{Summary of the main experimental results.}
\textbf{(a)} Few-shot adaptation across calibration sizes.
\textbf{(b)} Cross-dataset transfer across target-data fractions.
\textbf{(c)} Stage-1-only gains over matched scratch training.
\textbf{(d)} Stage-1 input ablation with Full, Coarse-only, and Neural-only variants.
\textbf{(e)} Representative trajectory reconstructions from Vel-Int and AMRT.}
    \label{fig:exp}
\end{figure}

\paragraph{Cross-dataset pretraining selectively improves target-data efficiency.}
The strongest transfer gains are observed on U-Limb. With Stage-1-only adaptation, both J$\rightarrow$U and W$\rightarrow$U outperform scratch training at every target-data fraction and across all split-fraction comparisons. With only 10\% of the U-Limb training data, they achieve PCC of 0.454 and 0.371, exceeding 0.192 obtained by scratch training with the full target data. This suggests that cross-dataset pretraining captures reusable structure that is difficult to learn from limited target data. However, these gains are not uniform across source-target pairs, motivating a closer examination of what determines transferability.

\paragraph{AMRT transfers both motion tendencies and reusable neural representations.}
Successful transfer does not necessarily require the source-domain motion prediction to remain directly valid. For W$\rightarrow$U, the frozen Stage-0 decoder already achieves a PCC of 0.310 without target-domain parameter updates, indicating that the learned motion tendency remains useful in the target domain. In contrast, J$\rightarrow$U has near-zero calibrated zero-update Stage-0 PCC, yet reaches an average PCC of 0.432 by adapting only Stage-1. Thus, even when the source motion output is poorly calibrated to the target, the frozen source stack can retain information that remains reusable after refinement.

To identify which information Stage-1 reuses during transfer, we ablate its two refinement inputs in Eq.~\ref{fuse}: the neural feature $\mathbf{h}^{(1)}_{t}$, derived from the frozen neural representation, and the coarse-trajectory embedding $\mathbf{h}^{p}_{t}$. \textbf{Coarse-only} removes $\mathbf{h}^{(1)}_{t}$, whereas \textbf{Neural-only} removes $\mathbf{h}^{p}_{t}$. As shown in Figure~\ref{fig:exp}(d), W$\rightarrow$U is largely coarse-dominant: coarse-only nearly matches the full model ($+0.23$ vs.\ $+0.24$), whereas neural-only yields only a small and less consistent gain ($+0.06$). In contrast, both inputs contribute to J$\rightarrow$U, with coarse-only and neural-only each improving over scratch ($+0.18$ and $+0.14$) and the full model achieving the largest gain ($+0.27$). These results suggest that W$\rightarrow$U mainly reuses source-derived motion information, while J$\rightarrow$U benefits from complementary coarse motion and target-readable neural representations. Together, these results show that AMRT transfers both motion information and reusable neural representations across datasets.

\paragraph{Transfer is strongly directional.}
Joystick-Track provides the strongest initialization for U-Limb despite differences in recording modality and control space. The channel adapter enables shared decoding across heterogeneous electrode configurations without requiring one-to-one channel correspondence. Successful J$\rightarrow$U transfer indicates that motion-related representations learned from iEEG can remain useful after alignment to EEG, highlighting the potential for cross-modal transfer. In contrast, neither EEG source transfers effectively to Joystick-Track. This asymmetry is consistent with differences in neural sampling and movement paradigms. EEG captures spatially mixed cortical activity, whereas ECoG provides more localized measurements with patient-specific electrode coverage (\citep{BURLE2015210, dubey2019cortical}). Joystick-Track involves constrained cursor tracking, whereas U-Limb spans diverse upper-limb movements. Direct trajectory-template transfer also yields higher PCC for J$\rightarrow$U than U$\rightarrow$J ($-0.16$ vs. $-0.41$), suggesting directional differences in motion-pattern compatibility. Together, these findings highlight the potential roles of neural representation and motion compatibility in cross-dataset transfer.

\subsection{Motion-to-Trajectory Decoding}
\label{sec:motion_to_trajectory}
Having evaluated \model{} under increasingly challenging generalization settings, from unseen trials to unseen subjects and datasets, we next examine which reconstruction formulation gives rise to these gains. Under the same model architecture, training configuration, and data splits, we compare \textsc{Direct-Pos}, which directly regresses trajectories; \textsc{Vel-Int}, which predicts velocity and reconstructs trajectories by temporal integration; \textsc{Pos-Vel}, which jointly predicts position and velocity while using the predicted position for reconstruction; and our full formulation. The three variants share the same Stage-0 architecture and differ only in their prediction objective, while the comparison between \textsc{Vel-Int} and our full model isolates the contribution of trajectory refinement. We evaluate all variants under the same \texttt{ctri-mix}, \texttt{ctri}, and \texttt{csub} protocols used above.
\input{tables/objvt}

Table~\ref{tab:motion_to_trajectory} reveals a clear interaction between reconstruction strategy and generalization regime. Under subject-specific \texttt{ctri}, direct position regression remains competitive, showing that absolute trajectories can be learned effectively when the subject-specific movement space is fixed. AMRT is not designed to dominate this setting; its advantage emerges as the decoding distribution shifts across subjects and recording conditions. Under \texttt{ctri-mix} and particularly \texttt{csub}, AMRT consistently achieves stronger trajectory reconstruction, with clearer gains on U-Limb and Joystick-Track. The weaker performance of \textsc{Pos-Vel} further shows that auxiliary velocity supervision alone does not explain these gains. Using velocity as an intermediate representation emphasizes local motion trends that are less tied to workspace, providing a more transferable basis for trajectory reconstruction.

Comparing \textsc{Vel-Int} with AMRT isolates the effect of trajectory refinement. While velocity captures local motion well, direct integration accumulates errors over time. Stage-1 mitigates this drift by refining the trajectory over longer horizons. This improvement is further supported by positional RMSE (Table \ref{tab:motion_to_trajectory_RMSE}), where AMRT consistently reduces reconstruction errors relative to Vel-Int across all evaluation settings. As illustrated in Figure~\ref{fig:exp}(e), \textsc{Vel-Int} follows the overall movement but gradually deviates from the ground truth, whereas AMRT corrects this drift while preserving the motion pattern. Together with Table~\ref{tab:motion_to_trajectory}, these results support the two-stage design: velocity anchors local motion, while trajectory refinement improves long-horizon consistency.

We extend the comparison of reconstruction strategies to cross-dataset transfer under Full-FT, covering all six transfer directions and four target-data fractions (Appendix~\ref{app:objvt-cdata}). Across the 24 settings, AMRT achieves the highest mean PCC in 21 and the lowest mean RMSE in 23, with consistent advantages on both metrics when U-Limb serves as the target. These results further demonstrate the effectiveness of AMRT's velocity-anchored, trajectory-refinement design for generalization across EEG/iEEG datasets.

%% file: tables/ctri-mix.tex

\begin{table}[b]
\centering
\footnotesize
\setlength{\tabcolsep}{1pt}
\caption{
Comparison of baselines and {\model}. Results are trajectory PCC. For Joystick-Track under \texttt{ctri-mix},
JT-common uses the channel subset shared across subjects, enabling matched comparison with fixed-channel baselines, while Joystick-Track uses all available electrodes.}
\label{tab:unseen_trial}
\resizebox{\linewidth}{!}{
\begin{tabular}{lcccccc}
\toprule
& \multicolumn{4}{c}{\texttt{ctri-mix}}
& \multicolumn{2}{c}{\texttt{ctri}} \\
\cmidrule(lr){2-5} \cmidrule(l){6-7}
& WAY-EEG-GAL
& \makecell{JT-common}
& \makecell{Joystick-Track}
& U-Limb
& WAY-EEG-GAL
& Joystick-Track \\
\midrule

M3T-Attention
& 0.813$\pm$0.036
& 0.464$\pm$0.097
& --
& 0.256$\pm$0.041
& 0.869$\pm$0.058
& 0.575$\pm$0.128 \\

Copilot-Assisted
& 0.674$\pm$0.022
& 0.378$\pm$0.084
& --
& 0.314$\pm$0.025
& 0.785$\pm$0.083
& 0.465$\pm$0.303 \\

TBR
& 0.248$\pm$0.053
& --
& 0.012$\pm$0.051
& 0.033$\pm$0.028
& 0.464$\pm$0.096
& 0.037$\pm$0.035 \\

\model{}
& \textbf{0.837$\pm$0.045}
& \textbf{0.707$\pm$0.037}***
& \textbf{0.787$\pm$0.064}***
& \textbf{0.530$\pm$0.029}***
& \textbf{0.914$\pm$0.028}*
& \textbf{0.834$\pm$0.107}** \\

\bottomrule
\end{tabular}
}
\end{table}

%% file: tables/csub.tex
\begin{table}[t]
\centering
\small
\caption{Comparison of baselines, the full AMRT, and the channel-adapter ablation (AMRT w/o CA), while AMRT w/o CA-res/si removes the learnable adapter residual for WAY-EEG-GAL and U-Limb, and spatial information for Joystick-Track. Best results are shown in bold. Significance markers are computed against the best baseline.}
\label{tab:unseen_sub}
\begin{tabular}{lcccc} 
\toprule
& WAY-EEG-GAL
& U-Limb
& Joystick-Track & JT-common \\
\midrule
M3T-Attention   & 0.785$\pm$0.128  & 0.160$\pm$0.061 & -- &0.334$\pm$0.110 \\
Copilot-Assisted  & 0.539$\pm$0.159 & 0.085$\pm$0.078 & -- &0.094$\pm$0.079 \\
TBR           & 0.263$\pm$0.100  & 0.033$\pm$0.028 & 0.013$\pm$0.041 & --  \\
AMRT w/o CA   & 0.789$\pm$0.103 & 0.360$\pm$0.105 & 0.646$\pm$0.138 & 0.673$\pm$0.107 \\
AMRT w/o CA-res/si & 0.824$\pm$0.068 & 0.372$\pm$0.114 & 0.680$\pm$0.116 & 0.679$\pm$0.092
\\
{\model}    & \textbf{0.841$\pm$0.055} & \textbf{0.430$\pm$0.073}*** & \textbf{0.719$\pm$0.109}*** & \textbf{0.687$\pm$0.125}***\\
\bottomrule
\end{tabular}
\end{table}

%% file: tables/objvt.tex
\begin{table}[t]
\centering
\footnotesize
\caption{
Trajectory PCC under different motion-to-trajectory reconstruction strategies. The best results are highlighted in bold.
}
\label{tab:motion_to_trajectory}
\setlength{\tabcolsep}{5pt}
\begin{tabular}{llcccc}
\toprule
Setting & Dataset
& \textsc{Direct-Pos}
& \textsc{Pos-Vel}
& \textsc{Vel-Int}
& {\model} \\
\midrule

\multirow{2}{*}{\texttt{ctri}}
& W & 0.934$\pm$0.022 & \textbf{0.935$\pm$0.024} & 0.886$\pm$0.040 & 0.914$\pm$0.028 \\
& J & 0.849$\pm$0.065 & \textbf{0.860$\pm$0.050} & 0.823$\pm$0.114 & 0.834$\pm$0.107 \\
\midrule

\multirow{3}{*}{\texttt{ctri-mix}}
& W & 0.831$\pm$0.020 & 0.831$\pm$0.020 & 0.785$\pm$0.038 & \textbf{0.837$\pm$0.045} \\
& U & 0.259$\pm$0.053 & 0.187$\pm$0.061 & 0.471$\pm$0.020 & \textbf{0.530$\pm$0.029}*** \\
& J & 0.702$\pm$0.018 & 0.694$\pm$0.020 & 0.770$\pm$0.076 & \textbf{0.787$\pm$0.064} \\
\midrule

\multirow{3}{*}{\texttt{csub}}
& W & 0.825$\pm$0.066 & 0.836$\pm$0.062 & 0.800$\pm$0.076 & \textbf{0.841$\pm$0.055} \\
& U & 0.268$\pm$0.120 & 0.282$\pm$0.123 & 0.369$\pm$0.107 & \textbf{0.430$\pm$0.073}* \\
& J & 0.361$\pm$0.060 & 0.357$\pm$0.052 & 0.689$\pm$0.132 & \textbf{0.719$\pm$0.109} \\

\bottomrule
\end{tabular}
\end{table}

%% file: paper/conclusion.tex
\section{conclusion}
We introduce \model{}, a channel-adaptive framework for generalizable continuous neural decoding. It combines a neuroscience-inspired two-stage formulation for motion anchoring and trajectory refinement with a shared decoding framework for heterogeneous neural recordings. Across a hierarchical evaluation from unseen trials and subjects to cross-dataset transfer, AMRT learns reusable neural representations and maintains strong decoding performance under increasingly challenging domain shifts. These results highlight the potential of combining transferable local motion with long-horizon trajectory modeling for generalizable continuous neural control. Future work can extend this framework to motor behaviors and larger-scale heterogeneous neural data.

%% file: paper/statement.tex
\section{AI Use Statement}
Generative AI tools were used to assist with writing, literature discovery, and research implementation. For writing, AI was used to revise and polish English expressions, improve clarity and conciseness, and refine the presentation of sections, captions, and experimental descriptions. For retrieval and discovery, AI was used to help identify potentially relevant prior work and organize related literature, with references subsequently checked by the authors. For research execution, AI was primarily used as a coding assistant to implement and debug experimental scripts and improve data processing and visualization code. All methodological decisions, experimental design, result interpretation, and final manuscript content were determined and verified by the authors.

\section{ethics statement}
We use only publicly available, de-identified neural datasets and conduct no new human-subject data collection. All recordings were collected under the ethics and consent procedures of the original studies. We follow the corresponding data-use requirements and report no additional identifiable participant information.

\section{reproducibility statement}
To facilitate reproducibility, we provide detailed dataset descriptions, preprocessing procedures, and evaluation protocols in Appendix~\ref{app:dataset}, together with the model architecture, optimization settings, and training configurations in Appendix~\ref{app:imply}. 

%% file: appendix/related.tex
\section{related work}
\subsection{Continuous Neural Decoding of Movement Kinematics}
Early continuous movement decoders relied on hand-crafted neural features and classical regressors. Linear regression and partial least squares mapped spectral or slow-cortical-potential features to limb kinematics \citep{khayyam2024comparison}, often combined with Kalman filtering for temporal smoothing \citep{cimarosto2024decoding, khoshkavandi2025hand}, while Gaussian processes and gradient-boosted trees \citep{tian2025interpretable} provided nonlinear alternatives. Hybrid approaches further replaced parts of state-space models with neural networks, such as learned transition dynamics \citep{khoshkavandi2025hand} or CNN-based EEG denoising \citep{khoshkavandi2025decoding}. More recent methods learn spatial-temporal representations directly from neural signals. CNN-GRU and CNN-LSTM architectures have improved hand, wrist, and lower-limb decoding over classical or purely convolutional models \citep{khayyam2024comparison, jain2023eeg, korik2025decoding}, with later extensions incorporating interpretable spectral-spatial weighting \citep{borra2023decoding}, lightweight preprocessing and correlation-aware objectives \citep{giri2026neurokinect}, cortical-source features \citep{jain2023eeg}, and attention for fine-grained finger-joint reconstruction \citep{tang2024reconstruction}. Despite these advances, most methods still rely on fixed convolutional receptive fields or sequential recurrent dynamics for temporal modeling.

\subsection{Attention-Based Temporal Modeling}
Continuous movement depends on neural dynamics across multiple temporal scales. Recurrent models capture temporal context but are sequential and difficult to parallelize \citep{wang2023mtrt}, while temporal convolutions are limited by fixed receptive fields \citep{kwon2026e2t}. Transformer-based methods address these limitations through self-attention, enabling sequence-level trajectory reconstruction \citep{wang2023mtrt}, frame-wise velocity decoding \citep{kwon2026e2t}, multi-scale temporal attention \citep{zhu2026m3t}, and spectral-domain attention \citep{fu2026sand}. Related work has also extended Transformers to jointly decode heterogeneous kinetic and kinematic variables \citep{dangi2026towards} and to synthesize whole-body motion from EEG \citep{peng2026eeg2motion}. However, attention alone is not always superior: under controlled comparisons, models retaining temporal convolutions can outperform attention-only variants \citep{ohsaki2026adaptation}. This suggests that local neural dynamics and long-range context are complementary, although most existing methods still combine them within a single pathway for direct position, velocity, or trajectory regression.

\subsection{Generalization Across Subjects}
Generalization remains challenging because neural recordings vary across subjects, electrode layouts, and acquisition settings. \citet{fu2026eeg} improves unseen-subject gait decoding through phase-aware EEG-motion alignment and domain-aware decoding, while SAND enhances cross-subject robustness by combining low-frequency spectral representations with temporal attention \citep{fu2026sand}. However, these methods largely operate with fixed channel organizations and task-specific kinematic spaces, without explicitly addressing heterogeneous electrode layouts or transferable motion representations across recording modalities and movement paradigms. AMRT addresses this gap through a channel-adaptive neural interface and a velocity-anchored, neural-conditioned trajectory refinement framework for generalization across subjects and datasets.

%% file: appendix/baseline.tex
\section{baseline}
We compare AMRT against three recent EEG-based continuous movement decoding methods: \textsc{Copilot}, \textsc{TBR}, and \textsc{M3T-Attention}. We select these baselines based on three criteria: they support continuous kinematic or trajectory regression, are evaluated on publicly available datasets rather than exclusively on proprietary recordings, and provide open-source implementations that allow faithful reproduction and adaptation to our evaluation protocol. Together, they represent recent approaches based on multi-scale convolutional modeling, Transformer-based temporal regression, and multi-scale temporal attention.

\textbf{Copilot} \citep{li2026copilot} employs a hybrid convolution-attention architecture for EEG-based hand-motion decoding. Its EEG decoder extracts temporal patterns at multiple receptive-field scales, applies squeeze-and-excitation modules to adaptively reweight channel information, and uses multi-head self-attention to capture long-range temporal dependencies. The resulting representation is directly mapped to continuous hand kinematics.

\textbf{TBR} \citep{dangi2026towards} formulates movement decoding as Transformer-based temporal regression over EEG-derived event-related desynchronization/synchronization (ERDS) features. A Transformer encoder models temporal dependencies within the feature sequence and jointly predicts multiple kinetic and kinematic movement variables. We use its regression architecture as a representative feature-based Transformer baseline for continuous decoding.

\textbf{M3T-Attention} \citep{zhu2026m3t} is a multi-level, multi-scale temporal attention model designed for EEG-based hand-trajectory decoding. It extracts temporal representations at multiple scales and integrates them through cross-scale attention, allowing the model to capture movement-related neural dynamics over different temporal ranges before predicting continuous hand kinematics.

%% file: appendix/dataset.tex
\section{Dataset Details}
\label{app:dataset}
\input{tables/dataset}
This appendix provides additional details on the three datasets summarized in Section~\ref{sec:setup}, including their recording setups, movement paradigms, data organization, and preprocessing procedures. A summary of their key characteristics is provided in Table~\ref{tab:datasets}.

\subsection{WAY-EEG-GAL}
WAY-EEG-GAL \citep{luciw2014multi} contains synchronized 32-channel scalp EEG and kinematic recordings acquired during grasp-and-lift movements. Each trial involves reaching toward an object, grasping and lifting it, and returning the hand to the resting position, under varying object weights and surface conditions. We decode the three-dimensional wrist trajectory, corresponding to the fourth position marker in the provided kinematic stream.
For preprocessing, each \texttt{WS\_P*\_S*.mat} file is unpacked into individual trials, with EEG and kinematic signals aligned at 500~Hz. The prediction target is represented as $[\dot{x},\dot{y},\dot{z},x,y,z]$, where $(x,y,z)$ denotes the wrist position and velocity is computed causally using a third-order Savitzky-Golay derivative with a 31-sample trailing window. Following recent continuous trajectory decoding studies \citep{giri2026neurokinect,zhu2026m3t}, we additionally apply a 0.5-12 Hz band-pass filter to the EEG signals.

\subsection{U-Limb}
U-Limb \citep{averta2021u} provides synchronized high-density EEG and motion-capture recordings during a diverse set of upper-limb movements. The protocol comprises 30 movement tasks, each performed repeatedly. We use all 128 EEG channels and decode the three-dimensional trajectory of the hand dorsum. Following the dataset-provided marker definitions and rigid-pose procedure, we recover the hand reference frame and define the hand-dorsum position as the centroid of the four associated markers.
We first apply basic quality control to retain valid trials. Trials are required to contain all four hand-dorsum markers, and those exhibiting implausible hand speeds above 3 m/s, typically caused by marker-tracking discontinuities, are removed. We further retain only participants with at least 20 valid trials to ensure sufficient data for training and evaluation.
Kinematics are sampled at 100 Hz and temporally aligned with the 500-Hz EEG recordings. We retain the hand-dorsum position $(x,y,z)$ and compute velocity using a causal third-order Savitzky--Golay derivative with a seven-sample trailing window at 100~Hz. The resulting prediction target is represented as $[\dot{x},\dot{y},\dot{z},x,y,z]$.

\subsection{Joystick-Track}
Joystick-Track \citep{miller2019library, schalk2007decoding} is an intracranial ECoG dataset, comprising recordings from four participants performing continuous two-dimensional cursor tracking along a circular target trajectory. The number of available ECoG channels varies across participants. The original task consists of repeated circular-tracking epochs, whereas the released data are stored as continuous recordings without explicit trial annotations.
We downsample the ECoG signals from 1 kHz to 500 Hz and the cursor coordinates are resampled to 100 Hz. The decoding target is represented as $[\dot{x},\dot{y},x,y]$, where $(x,y)$ denotes the observed cursor position and velocity is computed using a causal third-order Savitzky-Golay derivative with a seven-sample trailing window at 100 Hz. For trial-wise evaluation, we recover complete circular-tracking trials by unwrapping the target phase and discard incomplete boundary segments and duration outliers.

%% file: tables/dataset.tex
\begin{table}[t]
\centering
\small
\setlength{\tabcolsep}{3pt}
\caption{Summary of the datasets used in our experiments.}
\label{tab:datasets}
\begin{tabular}{lccccccc}
\toprule
Dataset & Modality & Task & Subjects & Channels & Trials  & Mean duration \\
\midrule
WAY-EEG-GAL
& EEG
& 6 conditions grasp-and-lift
& 12
& 32
& 3,528 
& 9.2 s
\\
U-Limb
& EEG
& 30 movement tasks
& 20
& 128
& 764
& 12.6 s
\\
Joystick-Track
& ECoG
& Circular joystick tracking
& 4
& 48-64
& 150
& 6.3s
\\
\bottomrule
\end{tabular}
\end{table}

%% file: appendix/model.tex
\section{Implementation Details}
\label{app:imply}
\subsection{Model Configuration}
Unless otherwise specified, we use the same model configuration across datasets, with only modality-dependent input handling and output dimensionality adapted to each setting. The encoder token rate, input context length, prediction horizon, and stride are configurable temporal hyperparameters. In our default setting, recordings are represented at $100$ Hz, with a $5$-s context ($H=500$), a $1$-s prediction horizon ($L=100$), and a stride of $0.5$s ($S=50$ tokens).

\input{tables/model-config}

\paragraph{Encoder.}
The channel adapter uses a hidden dimension of $128$, $4$ learned spatial queries, and $8$ attention heads. 
Each electrode token contains $P$ raw neural samples,
with $P=5$ by default, corresponding to a neural sampling rate of 500 Hz and a control rate of 100 Hz.
The waveform encoder consists of LayerNorm, a $5\rightarrow128$ linear projection, GELU, dropout, and a $128\rightarrow128$ linear layer. Local and global electrode coordinates are encoded by separate two-layer MLPs with dimensions $3\rightarrow64\rightarrow128$. The four spatial-query outputs are concatenated and projected to $128$ dimensions followed by GELU.

For a batch of size $B$, the adapter takes $\mathbf X\in\mathbb R^{B\times H\times C_{\mathrm{pad}}\times P}$, coordinates $\mathbf R\in\mathbb R^{B\times C_{\mathrm{pad}}\times3}$, and a validity mask $\mathbf m\in\{0,1\}^{B\times C_{\mathrm{pad}}}$, with at least one valid channel per sample and $P=5$ by default. Let $\sigma$ denote GELU and $\mathcal D$ dropout with rate 0.15; affine-layer biases are implicit. Suppressing the batch index, let $m_c$ indicate channel validity. Local geometry is computed as
{\footnotesize
\begin{align*}
n_s&=\sum_{c=1}^{C_{\mathrm{pad}}}m_c,\qquad
\boldsymbol{\mu}_s=\frac{1}{n_s}\sum_c m_c\mathbf r_{s,c},\\
\rho_s&=\left[\frac{1}{n_s}\sum_c m_c
\|\mathbf r_{s,c}-\boldsymbol{\mu}_s\|_2^2\right]^{1/2}, 
\widetilde{\mathbf r}_{s,c}=\frac{m_c(\mathbf r_{s,c}-\boldsymbol{\mu}_s)}
{\max(\rho_s,10^{-4})}.
\end{align*}
}

Invalid channels contribute neither to the center nor to the RMS radius, so additional masked padding leaves valid coordinates unchanged. The shared electrode encoders then compute
{\footnotesize
\begin{align*}
g_x(\mathbf x)
 &=\sigma\!\left(W_{x2}\mathcal D\!\left[
 \sigma(W_{x1}\operatorname{LN}_{P}(\mathbf x))\right]\right),
 & W_{x1}&\in\mathbb R^{128\times P},\quad
 W_{x2}\in\mathbb R^{128\times128},\\
g_j(\mathbf r)
 &=W_{j2}\sigma(W_{j1}\mathbf r),\quad j\in\{\ell,g\},
 & W_{j1}&\in\mathbb R^{64\times3},\quad
 W_{j2}\in\mathbb R^{128\times64},\\
\mathbf E_{t,c}
 &=m_c\left[g_x(\mathbf X_{t,c})
 +g_{\ell}(\widetilde{\mathbf r}_{s,c})
 +g_g(\mathbf r_{s,c}/\kappa)\right],
 & \mathbf E_t&\in\mathbb R^{C_{\mathrm{pad}}\times128},
 \quad\kappa=100.
\end{align*}
}
The coordinate embeddings are broadcast over the $H$ control steps. With learned queries $\mathbf S\in\mathbb R^{4\times128}$, spatial pooling uses $Q=\mathbf S$ and $K=V=\mathbf E_t$. For each of eight heads ($d_h=16$),
{
\footnotesize
\begin{align*}
\mathbf O_t^{(a)}
 &=\mathcal D\!\left[\operatorname{softmax}_{c}\!\left(
 \frac{(\mathbf S W_Q^{(a)})(\mathbf E_tW_K^{(a)})^\top}{4}
 +\mathbf M_c\right)\right]\mathbf E_tW_V^{(a)},\\
\mathbf O_t
 &=\operatorname{Concat}_{a=1}^{8}(\mathbf O_t^{(a)})W_O
 \in\mathbb R^{4\times128},\\
\mathbf u_t
 &=\mathcal D\!\left[\sigma\!\left(
 W_f\operatorname{LN}_{512}(\operatorname{vec}\mathbf O_t)
 \right)\right]\in\mathbb R^{128},
 \qquad W_f\in\mathbb R^{128\times512}.
\end{align*}
}
Here $W_Q^{(a)},W_K^{(a)},W_V^{(a)}\in\mathbb R^{128\times16}$, $W_O\in\mathbb R^{128\times128}$, and each row of $\mathbf M_c\in\mathbb R^{4\times C_{\mathrm{pad}}}$ is zero for valid channels and $-\infty$ otherwise, giving invalid channels zero attention weight. Pooling is independent at each control step. The global layout branch computes
{
\footnotesize
\begin{align*}
\eta_s
 &=\left[\frac{1}{3n_s}\sum_c m_c
 \|\mathbf r_{s,c}-\boldsymbol\mu_s\|_2^2\right]^{1/2}, 
\mathbf a_s=[\boldsymbol\mu_s/\kappa;\eta_s/\kappa;\log(n_s)/5]
 \in\mathbb R^5,\\
\mathbf h_t
 &=\mathcal D\!\left[\sigma\!\left(
 \operatorname{LN}_{128}\!\left[
 \mathbf u_t+W_{g2}^{\mathrm{ctx}}
 \sigma(W_{g1}^{\mathrm{ctx}}\mathbf a_s)\right]
 \right)\right]\in\mathbb R^{128},
\end{align*}
}
where $W_{g1}^{\mathrm{ctx}}\in\mathbb R^{128\times5}$ and $W_{g2}^{\mathrm{ctx}}\in\mathbb R^{128\times128}$. Thus the adapter returns $B\times H\times128$ tokens regardless of the number of valid electrodes. All electrode encoders and attention parameters are shared across channels and recordings.

The temporal encoder uses one bidirectional LSTM layer with hidden size $64$ per direction, followed by four multi-horizon temporal blocks. Each block operates at hidden dimension $128$ and contains depthwise temporal convolutions with receptive fields of $5$, $25$, $51$, and $101$ tokens. The global branch uses one learned query with $8$-head attention. The feed-forward expansion factor is $4$, and the maximum supported context length is $4096$ tokens.

\paragraph{Decoder.}
The Stage-0 decoder uses $L$ learned output queries, hidden dimension $128$, $8$-head query self-attention, and $8$-head cross-attention to the encoder memory. The output head is LayerNorm--Linear--GELU--Dropout--Linear. The decoder predicts three velocity dimensions $(v_x,v_y,v_z)$; for two-dimensional targets, the third dimension is omitted.
The decoder has one self-/cross-attention pair per stage. For stage $s$, let $\mathbf Q^{(s)}\in\mathbb R^{L\times128}$ be the learned queries, broadcast across the batch, and let $\mathbf Z\in\mathbb R^{B\times H\times128}$ be the current sample's encoder memory. The complete decoder path is 
{
\footnotesize
\begin{align*}
\mathbf U^{(s)}
 &=\operatorname{MHA}_{\mathrm{self}}^{(s)}
 (\mathbf Q^{(s)},\mathbf Q^{(s)},\mathbf Q^{(s)}),\\
\mathbf C^{(s)}
 &=\operatorname{MHA}_{\mathrm{cross}}^{(s)}
 (\mathbf U^{(s)},\mathbf Z,\mathbf Z;\mathbf M_z),\\
\mathbf A^{(s)}
 &=\operatorname{LN}_a^{(s)}
 (\mathbf Q^{(s)}+\mathbf C^{(s)}),\\
\mathbf H^{(s)}
 &=W_2^{(s)}\mathcal D\!\left[\sigma\!\left(
 W_1^{(s)}\operatorname{LN}_f^{(s)}(\mathbf A^{(s)})
 \right)\right].
\end{align*}
}
All intermediate tensors have shape $B\times L\times128$, and $W_1^{(s)},W_2^{(s)}$ are $128\times128$ affine maps applied independently at each output position. Each MHA uses eight heads of width 16, with learned query, key, value, and output projections and attention dropout 0.15. There is no residual or normalization around self-attention; the cross-attention result is added to the original queries before LayerNorm. The final MLP starts with a separate LayerNorm and has no residual connection. Stage-0 subsequently applies the stated velocity output head to $\mathbf H^{(0)}$; Stage-1 uses the same decoder structure with separate parameters. Self-attention permits all $L\times L$ query interactions. Cross-attention permits every query to read every valid position in the supplied $H$-step memory: $M_{z,ij}=0$ for valid memory position $j$, and $-\infty$ otherwise. Neither operation uses a triangular temporal mask. All $L$ learned queries are active in the decoding path; invalid target elements are excluded from the loss. 

The Stage-1 refinement decoder uses the same hidden dimension ($128$), number of queries ($L$), and number of attention heads ($8$) as Stage-0. The coarse trajectory is projected from $D$ dimensions to $128$ dimensions, concatenated with the decoder state, and fused through a $256\rightarrow128$ projection. Separate linear heads predict the residual increment and sigmoid gate. Overlapping predictions are merged using weighted overlap-add. Unless otherwise specified, Stage-1 receives both encoder memory and the coarse trajectory. 

\subsection{Training Configuration}
Stage-0 is trained with AdamW using a learning rate of $2\times10^{-4}$ and weight decay $10^{-4}$. Stage-1 freezes Stage-0 and is optimized with AdamW using a learning rate of $10^{-3}$, weight decay $10^{-4}$. One complete trial is used per optimization step, while Stage-0 windows are materialized in mini-batches of $8$ for memory efficiency. 
For Stage-0, we apply dynamic per-axis reweighting to the masked Huber loss to balance decoding difficulty across motion axes. For each training mini-batch, we compute $e_d=\mathrm{MSE}_d/\max(\mathrm{Var}(v_d),\epsilon)$ over valid velocity targets and set $\alpha_d=\mathrm{clip}(e_d/\max(\bar e,\epsilon),0.5,2.0)$, where $\bar e$ averages $e_d$ across valid axes.

Stage-1 objective combines the trajectory reconstruction loss, gap-increment supervision loss, temporal smoothness penalty, and gate-utility loss. Their weights are $1$, $0.5$, $0.02$, and $0.2$, respectively; the gate-utility margin is set to $0.01$.
Let $\mathbf{b}_t$ denote the coarse Stage-0 trajectory and $\mathbf{y}_t$ the ground-truth trajectory. Stage-1 predicts a correction $\mathbf{c}_t$, yielding $\hat{\mathbf{y}}_t=\mathbf{b}_t+\mathbf{c}_t$.  
We optimize the refined trajectory with a SmoothL1 reconstruction loss, where $D$ indexes the kinematic control dimension:

\vspace{-1em}
{
\footnotesize
\begin{equation}
    \mathcal{L}_{\mathrm{traj}}
    =
    \frac{1}{|\mathcal{V}|}
    \sum_{(t,D)\in\mathcal{V}}
    \operatorname{SmoothL1}(\hat{y}_{t,D},y_{t,D}),
\end{equation}
and directly supervise the accumulated raw correction $\bar{\mathbf{c}}_t$
toward the residual gap $\mathbf{y}_t-\mathbf{b}_t$:
\begin{equation}
    \mathcal{L}_{\mathrm{gap}}
    =
    \frac{1}{|\mathcal{V}|}
    \sum_{(t,D)\in\mathcal{V}}
    \operatorname{SmoothL1}(\bar{c}_{t,D},y_{t,D}-b_{t,D}).
\end{equation}
}
\vspace{-1em}

To encourage temporally coherent corrections, we penalize consecutive changes,

\vspace{-1em}
{
\footnotesize
\begin{equation}
    \mathcal{L}_{\mathrm{smooth}}
    =
    \frac{1}{|\mathcal{V}_{\Delta}|}
    \sum_{(t,D)\in\mathcal{V}_{\Delta}}
    |c_{t,D}-c_{t-1,D}|.
\end{equation}
}
\vspace{-1em}

We additionally supervise the gate $g_{t,D}$ to activate only when the predicted correction improves the coarse trajectory. Specifically,

\vspace{-1em}
{
\footnotesize
\begin{equation}
    u_{t,D}
    =
    \mathbb{I}\!\left[
    |b_{t,D}+\bar{c}_{t,D}-y_{t,D}|+\delta
    <
    |b_{t,D}-y_{t,D}|
    \right],
    \qquad \delta=0.01,
\end{equation}
}
\vspace{-1em}

and define

\vspace{-2.5em}
{
\footnotesize
\begin{align}
    \mathcal{L}_{\mathrm{utility}}
    &=
    \frac{1}{|\mathcal{V}|}
    \sum_{(t,D)\in\mathcal{V}}
    \operatorname{BCE}(g_{t,D},u_{t,D}).
\end{align}
}
\vspace{-1.5em}

The complete Stage-1 objective is

\vspace{-1em}
{
\footnotesize
\begin{equation}
    \mathcal{L}_{\mathrm{Stage\text{-}1}}
    =
    \mathcal{L}_{\mathrm{traj}}
    +0.5\,\mathcal{L}_{\mathrm{gap}}
    +0.02\,\mathcal{L}_{\mathrm{smooth}}
    +0.2\,\mathcal{L}_{\mathrm{utility}}.
\end{equation}
}
\vspace{-1em}

\paragraph{Within-dataset and cross-subject training.}
For within-dataset experiments, both stages are trained on the corresponding training partition, with Stage-0 initialized from scratch and Stage-1 trained from the selected Stage-0 checkpoint. For subject-disjoint evaluation, Stage-0 is trained on source subjects and evaluated directly on the held-out target subject. In cross-subject experiments, for each held-out subject, few-shot adaptation draws $K\in{1,3,5,10}$ complete trials from a pool of 10 target-subject training trials under three different selection seeds, and results are averaged. Test trials are never used for adaptation, normalization, or checkpoint selection. In the full-adaptation protocol, Stage-0 is first fine-tuned for 15 epochs with a learning rate of $5\times10^{-5}$, followed by 15 epochs of Stage-1 refinement-head training with a learning rate of $10^{-5}$. In the Stage-1-only protocol, the transferred Stage-0 model remains frozen and only Stage-1 is fine-tuned for 15 epochs at $10^{-5}$. 

\paragraph{Cross-dataset transfer.}
Source models are trained on the full source training split. Target adaptation uses $f\in\{0,0.10,0.25,0.50,1.00\}$ of the target training trials. The $f=0$ setting denotes calibrated zero-update transfer: no model parameters are updated on the target domain, while velocity denormalization uses control statistics computed from the full target training split. Stage-0 uses a batch size of $16$ and AdamW with learning rate $2\times10^{-4}$. Two-source pretraining runs for $6{,}000$ optimization steps, while target-only training and transferred Stage-0 fine-tuning use $1{,}500$ steps. Stage-1 is trained for $20$ epochs using AdamW with learning rate $10^{-3}$, weight decay $10^{-4}$, and gradient clipping at $1.0$. Stage-0 windows are materialized in batches of $8$, while each refinement update uses one complete trial. 

\input{tables/train-config}


%% file: tables/model-config.tex
\begin{table}[b]
\centering
\small
\caption{Architecture hyperparameters of AMRT.}
\label{tab:arch_hparams}
\begin{tabular}{ll}
\toprule
\textbf{Component} & \textbf{Configuration} \\
\midrule
Hidden dimension & 128 \\
Dropout & 0.15 \\
Spatial queries & 4 \\
Attention heads & 8 \\
BiLSTM hidden size & 64 per direction \\
Temporal blocks & 4 \\
Temporal receptive fields & 5, 25, 51, 101 \\
Stage-0 output queries & $L$ \\
Stage-1 output queries & $L$ \\
Max context length & 4096 \\
\bottomrule
\end{tabular}
\end{table}

%% file: tables/train-config.tex
\begin{table}[h]
\centering
\small
\caption{Default training hyperparameters for cross-trial and cross-subject.}
\label{tab:train_hparams}
\begin{tabular}{ll}
\toprule
\textbf{Hyperparameters} & \textbf{Settings} \\
\midrule
Seed &42 \\
Stage-0 optimizer & AdamW \\
Stage-0 learning rate & $2\times10^{-4}$ \\
Stage-0 weight decay & $10^{-4}$ \\
Stage-0 batch size & 32(J), 128(others) \\
Stage-0 epochs & 100 \\
Stage-1 optimizer & AdamW \\
Stage-1 learning rate & $10^{-3}$ \\
Stage-1 weight decay & $10^{-4}$ \\
Stage-1 epochs & 150(J), 50(others) \\
Materialized Windows batch size & 8 \\
Increment loss weight & 0.5 \\
Smoothness loss weight & 0.02 \\
\bottomrule
\end{tabular}
\end{table}

%% file: appendix/experiment.tex
\section{Supplementary experiments}
\subsection{Additional Results}
\subsubsection{Additional Evaluation with Positional RMSE}
To complement trajectory PCC, which measures temporal correlation but does not capture positional magnitude errors, we additionally evaluate root mean squared error (RMSE) across all datasets and generalization settings. RMSE is computed between predicted and ground-truth relative trajectories over valid time steps and position axes, using the same test partitions as the PCC evaluation. As shown in Tables \ref{tab:main_rmse}-\ref{tab:motion_to_trajectory_RMSE}, AMRT achieves the lowest RMSE on U-Limb and Joystick-Track under ctri-mix and cross-subject evaluation, demonstrating improved positional accuracy in heterogeneous decoding settings. On WAY-EEG-GAL, most methods achieve relatively strong and comparable performance, potentially due to its highly stereotyped movement trajectories, as even an EEG-free trajectory template achieves high PCC and low RMSE, suggesting that shared movement patterns may contribute substantially to decoding performance. RMSE is reported in dataset-specific position units and compared only within each dataset.

\input{tables/RMSE}
\input{tables/objvt-RMSE}

\subsubsection{Additional Generalization Results}
\input{tables/few-shot-csub}
Table~\ref{tab:fewshot_full} reports the complete few-shot calibration results corresponding to the analysis in the main text. For each held-out subject, we start from the model trained under the subject-disjoint LOSO protocol and reserve 10 target-subject trials as a calibration pool, with all remaining trials used as a fixed test set. Zero-shot uses no calibration trials, while $K$-shot adaptation uses $K\in{1,3,5,10}$ trials from this pool before evaluation on the same test set. We compare Full-FT with Stage-1-only adaptation.
The full results further show that the preferred adaptation strategy changes with calibration size. Stage-1-only performs better in the lowest-shot settings, particularly on U-Limb and Joystick-Track, whereas full fine-tuning becomes more effective with additional supervision on WAY-EEG-GAL and U-Limb. On Joystick-Track, full fine-tuning remains below the zero-shot baseline across all shot counts, while Stage-1-only preserves or slightly improves performance.

\subsection{Cross-Dataset Transfer}
\label{app:cdata-app}
\subsubsection{Single-Source Transfer}
Table~\ref{tab:single_source_full} reports the complete single-source transfer results across all source-target pairs and target-data fractions, complementing the summary results in Section~\ref{sec:cross_dataset}. For each target fraction, we compare target-only training from scratch with Full-FT and Stage-1-only transfer from each source dataset.
Table~\ref{tab:stage1_input_ablation} provides the complete Stage-1 input ablation for the two successful transfer directions to U-Limb. Stage-1 receives two complementary inputs in the full model: (1) the neural memory produced by the frozen neural pathway, and (2) an embedding of the coarse trajectory reconstructed by Stage-0. Accordingly, \textbf{Coarse-only} removes the neural-memory input and refines the trajectory using only the coarse-trajectory embedding, whereas \textbf{Neural-only} removes the coarse-trajectory embedding and relies only on the neural memory. \textbf{Full} retains both inputs. These results provide the complete numerical counterpart to Figure~\ref{fig:exp}(d).
\input{tables/single-source}
\input{tables/S1-ablation}

\subsubsection{Dual-Source Transfer}
\label{app:exp-2source}
Table~\ref{tab:dual_source_transfer} and Figure~\ref{fig:exp_app}(a) report dual-source transfer, where each target is pretrained jointly on the other two datasets. The benefit remains strongly target-dependent. For U-Limb, both adaptation strategies consistently outperform scratch: Stage-1-only performs best with 10-50\% target data, while Full-FT becomes more effective at 100\%. For WAY-EEG-GAL, Stage-1-only is the most reliable strategy and achieves the best performance at all target-data fractions, reaching $0.831$ at 100\%. In contrast, neither adaptation strategy improves Joystick-Track over scratch. Full-FT remains slightly below scratch across fractions, while Stage-1-only degrades substantially, consistent with the weak transfer to Joystick-Track observed in the single-source experiments.

Comparing dual-source with single-source transfer further shows that adding a second source is not uniformly beneficial. For U-Limb, the best single-source models remain stronger than dual-source pretraining at every target-data fraction; in particular, J$\rightarrow$U with Stage-1-only achieves $0.454$ at 10\% target data, compared with $0.333$ for dual-source transfer. 
For WAY-EEG-GAL, dual-source pretraining provides mixed results compared with the best single-source transfer. Under Stage-1-only adaptation, combining U and J slightly outperforms J$\rightarrow$W at 25\% and 100\% target data (0.789 vs. 0.786 and 0.816 vs. 0.808), but performs worse at 10\% and 50\%. These results suggest that adding a second source does not consistently improve transfer performance.
For Joystick-Track, dual-source pretraining does not recover the degradation already observed with either EEG source. These results reinforce that transfer depends on source-target compatibility rather than source quantity alone: adding heterogeneous source data can improve transfer, but does not guarantee gains when the transferred representations are poorly matched to the target.

\input{tables/dual-source}

\subsubsection{Scaling with Heterogeneous Neural Data}
We further study the scaling behavior of joint continuous decoding as the amount of balanced heterogeneous neural data increases. For each dataset, complete trials are first split into training, validation, and test partitions at an $8{:}1{:}1$ ratio. The training and validation partitions are pooled across datasets, whereas each test partition is retained for dataset-specific evaluation. Across scaling settings, we vary only the composition and size of the pooled training set while keeping the validation and test partitions fixed. Checkpoints are selected on the pooled validation set using a dataset-size-weighted validation score.
Let $J$ denote the available training-set size of Joystick-Track, the smallest dataset. 
We construct balanced mixtures at four scales,
\[
(J/4,J/4,J/4),\;
(J/2,J/2,J/2),\;
(3J/4,3J/4,3J/4),\;
(J,J,J),
\]
where each tuple gives the numbers of Joystick-Track, U-Limb, and WAY-EEG-GAL training trials. Thus, training data increase uniformly across datasets without changing their relative contribution. The optimization budget scales proportionally with data size, while the architecture and evaluation protocol remain fixed. We repeat each setting over 5 random seeds and report the aggregated results.

As shown in Table~\ref{tab:trial_scaling}, increasing the balanced heterogeneous training set leads to an overall upward trend in decoding performance, with macro PCC increasing from $0.492$ at 90 trials to $0.595$ at 360 trials. Joystick-Track shows the clearest improvement with scale, while WAY-EEG-GAL improves at intermediate scales and U-Limb varies more across settings. Despite these dataset-specific fluctuations, the overall trend suggests that joint training may benefit from larger amounts of balanced heterogeneous neural data. These results suggest that scaling balanced heterogeneous neural data has the potential to improve shared continuous decoding.
\input{tables/scaling}

\subsubsection{Cross-dataset Comparison of Decoding Formulations}
\label{app:objvt-cdata}
We extend the reconstruction-strategy comparison in Sec. \ref{sec:motion_to_trajectory} to all six single-source transfer directions, evaluating AMRT, DIRECT-POS, POS-VEL, VEL-INT, and POS-RES under Full-FT, where POS-RES predicts trajectories in Stage-0 and applies the same Stage-1 residual refinement mechanism as AMRT. Following Sec. \ref{sec:cross_dataset}, we use identical data splits, random seeds, and target-training fractions (10\%, 25\%, 50\%, and 100\%). We report trajectory PCC and RMSE as mean $\pm$ standard deviation across runs (Table \ref{tab:cross_dataset_reconstruction}).

\input{tables/objvt-cdata}

The relative benefits of different reconstruction strategies vary across target datasets. For U-Limb, AMRT consistently achieves the best PCC and RMSE across both transfer directions and all data fractions, with particularly large PCC gains over alternative formulations. For WAY-EEG-GAL, AMRT generally improves PCC while consistently reducing RMSE. Notably, VEL-INT sometimes achieves comparable or higher PCC but substantially larger RMSE, indicating that velocity integration can preserve overall motion patterns while introducing considerable positional errors. For Joystick-Track, all reconstruction strategies struggle to achieve positive transfer over scratch training. Although AMRT generally achieves the best performance among them, it does not consistently outperform scratch, highlighting the challenges of transferring to heterogeneous iEEG recordings. To isolate the contribution of velocity anchoring, we further compare POS-RES, which applies the same Stage-1 refinement to direct trajectory predictions. Although POS-RES improves upon DIRECT-POS, AMRT consistently achieves superior performance, highlighting the generalization advantages of its velocity-to-trajectory design.

\subsection{Additional Model Analyses}
\label{app:model-abl}
\subsubsection{Model Ablations}
We further ablate the temporal encoder and action-chunk decoder under the \texttt{ctri-mix} protocol to isolate the contribution of the main architectural components. For the encoder, \textbf{w/o BiLSTM} removes the BiLSTM used to model temporal dependencies, while \textbf{w/o Multi-Horizon} removes the block that combines neural features from multiple temporal horizons. For the decoder, \textbf{Pooled Decoder} replaces the action-chunk decoder with a single globally pooled neural representation shared across all predicted future steps, removing step-specific access to the neural sequence.
We additionally evaluate three Stage-1 ablations using the same frozen pretrained Stage-0 model and trajectory-level supervision. \textbf{w/o Gate} removes the learned gate while retaining cumulative incremental correction. \textbf{w/o Gate-Increment} removes both mechanisms and directly predicts position corrections. These controls assess the contribution of the proposed correction mechanisms while controlling for the frozen front end, additional trainable capacity, and trajectory-level supervision.

\input{tables/model-ablation}

The encoder ablations show that both recurrent temporal modeling and multi-horizon aggregation contribute to decoding performance. Removing the BiLSTM reduces PCC most notably on U-Limb and Joystick-Track, while removing the multi-horizon block causes the largest overall degradation, particularly on U-Limb and WAY-EEG-GAL. Replacing the action-chunk decoder with temporal pooling also consistently reduces performance. These results indicate that AMRT benefits from both multi-scale temporal encoding and step-specific decoding of the motion chunk, rather than relying on a single globally pooled neural representation.

\begin{figure}[h]
    \centering
    \includegraphics[width=0.98\linewidth]{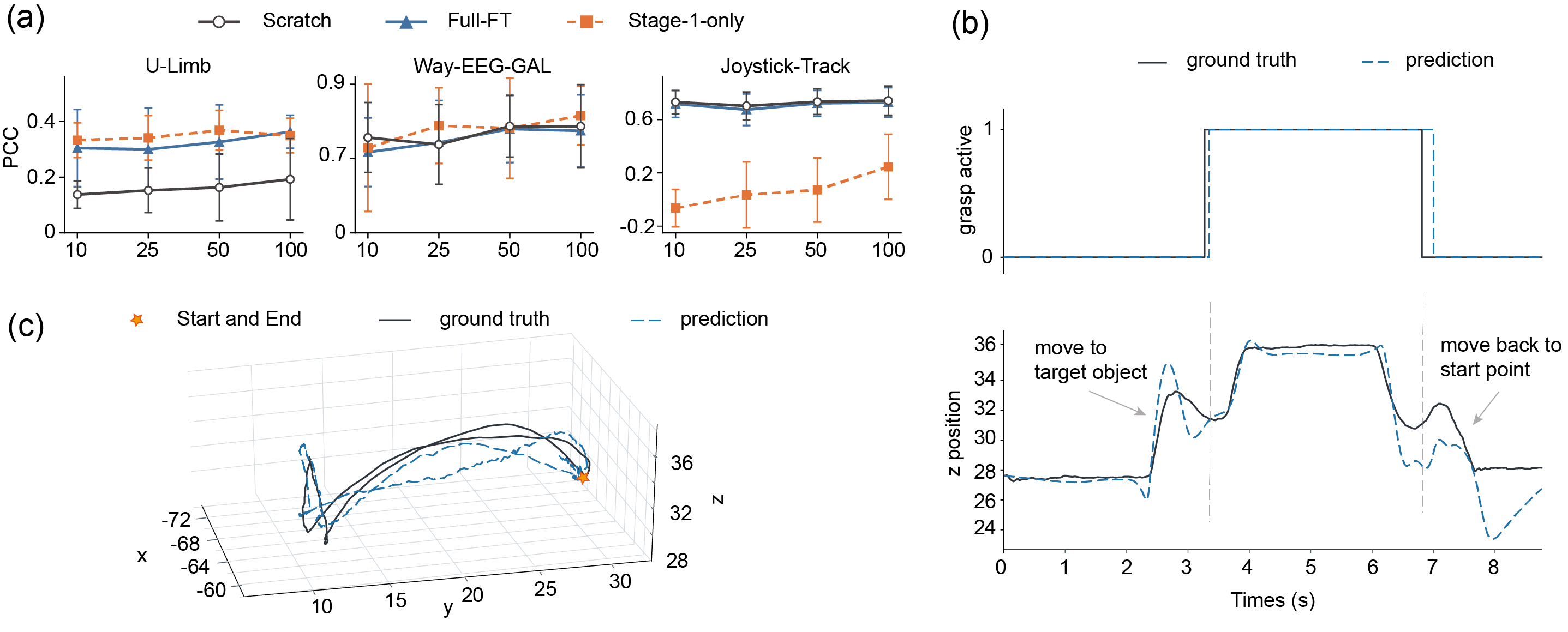}
    \caption{
    \textbf{Summary of supplementary experimental results.}
    \textbf{(a)} Dual-source cross-dataset transfer across target-data fractions.
    \textbf{(b)} Example of simultaneous trajectory and grasp-state decoding on WAY-EEG-GAL: ground-truth and predicted grasp state (top), and the corresponding $z$-axis trajectory (bottom).
    \textbf{(c)} Three-dimensional reconstruction of a representative WAY-EEG-GAL trajectory, comparing the prediction with the ground truth in the original movement space.
}
    \label{fig:exp_app}
\end{figure}

\subsubsection{grasp-state decoding}
WAY-EEG-GAL additionally provides grasp-state annotations, allowing us to examine whether the Stage-0 representation learned for continuous motion decoding also captures discrete motor-state information. We therefore evaluate binary grasp-state decoding (grasp vs.\ non-grasp) under the same \texttt{ctri}, \texttt{ctri-mix}, and \texttt{csub} protocols used for trajectory decoding. Specifically, the grasp classifier shares the neural encoder and Stage-0 action-chunk decoder with the motion prediction branch, and applies a lightweight binary classification head to the resulting action-chunk representation. The grasp head is optimized with binary cross-entropy loss. 
As shown in Table~\ref{tab:grasp_decoding}, grasp state remains consistently decodable across evaluation protocols, including the unseen-subject setting. This suggests that the Stage-0 action-chunk representation retains grasp-related motor-state information beyond continuous kinematics.
Figure~\ref{fig:exp_app}(b) provides a qualitative example of joint continuous and discrete motor decoding. We show the $z$-axis trajectory because vertical motion is directly related to the grasp-and-lift phase of the task. The alignment between grasp transitions and vertical movement further illustrates that the shared Stage-0 representation captures coordinated information about both motor state and continuous kinematics.

\input{tables/grasp}

\subsubsection{case study}
Figure~\ref{fig:exp_app}(c) provides a three-dimensional view of the representative trial analyzed along individual coordinate axes in Figure~\ref{fig:exp}(e). While the axis-wise plots show how Stage~1 corrects errors in each coordinate, the 3D visualization further shows that these corrections form a coherent spatial trajectory. The reconstructed path closely follows the overall geometry of the ground truth, including the major changes in movement direction and elevation. Residual deviations mainly occur around sharper local transitions, where the prediction remains smoother than the recorded trajectory.



%% file: tables/RMSE.tex
\begin{table}[b]
\centering
\footnotesize
\setlength{\tabcolsep}{1.5pt}
\caption{
Comparison of baselines and \model{}. Results are trajectory RMSE.}
\label{tab:main_rmse}
\resizebox{\linewidth}{!}{
\begin{tabular}{lccccc}
\toprule
& \multicolumn{3}{c}{\texttt{ctri-mix}}
& \multicolumn{2}{c}{\texttt{ctri}} \\
\cmidrule(lr){2-4} \cmidrule(l){5-6}
& WAY-EEG-GAL
& \makecell{Joystick-Track}
& U-Limb
& WAY-EEG-GAL
& Joystick-Track \\
\midrule

M3T-Attention
& \textbf{4.924$\pm$1.137}
& --
& 108.410$\pm$6.499 
& \textbf{3.088$\pm$0.659}
& 8745.303$\pm$418.109 \\

Copilot-Assisted
& 5.292$\pm$0.161
& --
& 96.721$\pm$3.891
& 4.344$\pm$0.902
& 9521.424$\pm$2389.939 \\

TBR
& 7.364$\pm$2.139
& 8719.876$\pm$1873.751 
& 98.770$\pm$18.371
& 6.515$\pm$0.591
& 8502.913$\pm$1823.509  \\

\model{}
& 5.237$\pm$1.159
& \textbf{7692.757$\pm$616.446}
& \textbf{80.926$\pm$3.760}
& 4.047$\pm$0.616
& \textbf{7422.698$\pm$604.229}\\

\bottomrule
\end{tabular}
}
\end{table}

\begin{table}[t]
\centering
\small
\caption{Comparison of baselines and \model{}. Results are trajectory RMSE.}
\label{tab:csub_rmse}
\begin{tabular}{lccc} 
\toprule
& WAY-EEG-GAL
& U-Limb
& Joystick-Track \\
\midrule
M3T-Attention  & \textbf{5.088$\pm$1.049} & 173.312$\pm$52.440 & -- \\
Copilot-Assisted  & 6.254$\pm$0.900 & 196.135$\pm$58.474 & -- \\
TBR           & 7.574$\pm$0.634   & 117.822$\pm$24.034 & 9559.497$\pm$393.377  \\
{\model}    & 5.809$\pm$0.565 & \textbf{79.106$\pm$13.723} & \textbf{8494.897$\pm$1492.847} \\
\bottomrule
\end{tabular}
\end{table}

%% file: tables/objvt-RMSE.tex
\begin{table}[t]
\centering
\footnotesize
\caption{
Trajectory RMSE under different motion-to-trajectory reconstruction strategies. The best results are highlighted in bold.
}
\label{tab:motion_to_trajectory_RMSE}
\setlength{\tabcolsep}{1.5pt}
\begin{tabular}{llcccc}
\toprule
Setting & Dataset
& \textsc{Direct-Pos}
& \textsc{Pos-Vel}
& \textsc{Vel-Int}
& {\model} \\
\midrule

\multirow{2}{*}{\texttt{ctri}}
& W & 2.330 $\pm$0.357 & \textbf{2.298$\pm$0.374} & 16.764$\pm$13.366 & 4.047 $\pm$0.616 \\
& J & 6075.707$\pm$326.454 & \textbf{5815.055$\pm$366.050} & 9028.251$\pm$1842.146 & 7422.698$\pm$604.229 \\
\midrule

\multirow{3}{*}{\texttt{ctri-mix}}
& W & 4.184$\pm$0.262 & \textbf{4.156$\pm$0.297} & 12.260$\pm$4.502 & 5.237$\pm$1.159 \\
& U & 115.118$\pm$1.935 & 111.472$\pm$1.297 & 85.979$\pm$4.331 & \textbf{80.926$\pm$3.760} \\
& J & \textbf{6686.755$\pm$253.153} & 6766.083$\pm$199.182 & 8591.401$\pm$1070.737 & 7692.757$\pm$616.446 \\
\midrule

\multirow{3}{*}{\texttt{csub}}
& W & 4.008$\pm$0.633 & \textbf{3.654$\pm$0.677} & 20.814$\pm$15.911 & 5.809$\pm$0.565 \\
& U & 129.277$\pm$43.680 & 122.090$\pm$27.688 & 87.462$\pm$16.057 & \textbf{79.106$\pm$13.723} \\
& J & 20412.376$\pm$1700.183 & 20363.037$\pm$1749.986 & 13465.264$\pm$1075.413 & \textbf{8494.897$\pm$1492.847} \\

\bottomrule
\end{tabular}
\end{table}

%% file: tables/few-shot-csub.tex
\begin{table}[h]
\centering
\caption{
Few-shot calibration on unseen subjects.
Results are trajectory PCC (mean$\pm$std) across held-out subjects.
The zero-shot result denotes performance without target-subject calibration. 
Bold indicates the better adaptation strategy at each shot count.
}
\label{tab:fewshot_full}
\footnotesize
\setlength{\tabcolsep}{3.5pt}
\begin{tabular}{llccccc}
\toprule
Dataset & Adaptation & zero-shot & 1-shot & 3-shot & 5-shot & 10-shot \\
\midrule

\multirow{2}{*}{WAY-EEG-GAL}
& Full-FT
& \multirow{2}{*}{0.837${\pm}$0.056}
& 0.819${\pm}$0.078
& \textbf{0.838${\pm}$0.073}
& \textbf{0.853${\pm}$0.061}
& \textbf{0.873${\pm}$0.030} \\
& Stage-1-only
&
& \textbf{0.822${\pm}$0.073}
& 0.838${\pm}$0.060
& 0.841${\pm}$0.058
& 0.849${\pm}$0.057 \\
\midrule

\multirow{2}{*}{U-Limb}
& Full-FT
& \multirow{2}{*}{0.427${\pm}$0.073}
& 0.381${\pm}$0.117
& 0.413${\pm}$0.113
& \textbf{0.476${\pm}$0.105}
& \textbf{0.525${\pm}$0.100} \\
& Stage-1-only
&
& \textbf{0.400${\pm}$0.090}
& \textbf{0.440${\pm}$0.084}
& 0.457${\pm}$0.086
& 0.476${\pm}$0.085 \\
\midrule

\multirow{2}{*}{Joystick-Track}
& Full-FT
& \multirow{2}{*}{0.710${\pm}$0.117}
& 0.579${\pm}$0.208
& 0.653${\pm}$0.121
& 0.633${\pm}$0.172
& 0.679${\pm}$0.132 \\
& Stage-1-only
&
& \textbf{0.714${\pm}$0.123}
& \textbf{0.710${\pm}$0.122}
& \textbf{0.710${\pm}$0.119}
& \textbf{0.722${\pm}$0.120} \\
\bottomrule
\end{tabular}
\end{table}

%% file: tables/single-source.tex
\begin{table}[h]
\centering
\caption{
Complete single-source cross-dataset transfer results.
Results are trajectory PCC.
Bold indicates the best result for each transfer direction and target-data fraction.
$J$, $U$, and $W$ denote Joystick-Track, U-Limb, and WAY-EEG-GAL.
}
\label{tab:single_source_full}
\small
\setlength{\tabcolsep}{5.0pt}
\begin{tabular}{llcccc}
\toprule
Transfer & Adaptation & 10\% & 25\% & 50\% & 100\% \\
\midrule

\multirow{3}{*}{$J\rightarrow U$}
& Scratch
& 0.137${\pm}$0.049
& 0.152${\pm}$0.080
& 0.163${\pm}$0.120
& 0.192${\pm}$0.146 \\
& Full-FT
& 0.337${\pm}$0.135
& 0.311${\pm}$0.139
& 0.411${\pm}$0.132
& 0.301${\pm}$0.051 \\
& Stage-1-only
& \textbf{0.454${\pm}$0.056}
& \textbf{0.434${\pm}$0.048}
& \textbf{0.426${\pm}$0.077}
& \textbf{0.413${\pm}$0.063} \\
\midrule

\multirow{3}{*}{$W\rightarrow U$}
& Scratch
& 0.137${\pm}$0.049
& 0.152${\pm}$0.080
& 0.163${\pm}$0.120
& 0.192${\pm}$0.146 \\
& Full-FT
& 0.264${\pm}$0.136
& 0.261${\pm}$0.183
& 0.376${\pm}$0.037
& 0.301${\pm}$0.148 \\
& Stage-1-only
& \textbf{0.371${\pm}$0.072}
& \textbf{0.387${\pm}$0.067}
& \textbf{0.419${\pm}$0.076}
& \textbf{0.427${\pm}$0.043} \\
\midrule

\multirow{3}{*}{$J\rightarrow W$}
& Scratch
& 0.757${\pm}$0.094
& 0.738${\pm}$0.107
& 0.787${\pm}$0.083
& 0.787${\pm}$0.112 \\
& Full-FT
& 0.741${\pm}$0.082
& 0.767${\pm}$0.095
& 0.775${\pm}$0.105
& 0.781${\pm}$0.099 \\
& Stage-1-only
& \textbf{0.787${\pm}$0.089}
& \textbf{0.786${\pm}$0.099}
& \textbf{0.805${\pm}$0.105}
& \textbf{0.808${\pm}$0.106} \\
\midrule

\multirow{3}{*}{$U\rightarrow W$}
& Scratch
& \textbf{0.757${\pm}$0.094}
& 0.738${\pm}$0.107
& \textbf{0.787${\pm}$0.083}
& 0.787${\pm}$0.112 \\
& Full-FT
& 0.686${\pm}$0.084
& 0.749${\pm}$0.109
& 0.780${\pm}$0.086
& \textbf{0.795${\pm}$0.077} \\
& Stage-1-only
& 0.747${\pm}$0.084
& \textbf{0.756${\pm}$0.066}
& 0.750${\pm}$0.085
& 0.760${\pm}$0.087 \\
\midrule

\multirow{3}{*}{$U\rightarrow J$}
& Scratch
& \textbf{0.730${\pm}$0.087}
& \textbf{0.702${\pm}$0.104}
& \textbf{0.733${\pm}$0.096}
& \textbf{0.742${\pm}$0.109} \\
& Full-FT
& 0.703${\pm}$0.116
& 0.690${\pm}$0.097
& 0.730${\pm}$0.093
& 0.727${\pm}$0.131 \\
& Stage-1-only
& 0.357${\pm}$0.226
& 0.484${\pm}$0.121
& 0.489${\pm}$0.104
& 0.582${\pm}$0.137 \\
\midrule

\multirow{3}{*}{$W\rightarrow J$}
& Scratch
& \textbf{0.730${\pm}$0.087}
& 0.702${\pm}$0.104
& \textbf{0.733${\pm}$0.096}
& \textbf{0.742${\pm}$0.109} \\
& Full-FT
& 0.725${\pm}$0.080
& \textbf{0.702${\pm}$0.119}
& 0.724${\pm}$0.105
& 0.729${\pm}$0.107 \\
& Stage-1-only
& -0.047${\pm}$0.224
& 0.084${\pm}$0.190
& 0.267${\pm}$0.117
& 0.320${\pm}$0.110 \\
\bottomrule
\end{tabular}
\end{table}

%% file: tables/S1-ablation.tex
\begin{table}[t]
\centering
\caption{
Stage-1 input ablation for single-source transfer to U-Limb under Stage-1-only adaptation. Results are trajectory PCC (mean$\pm$std); bold indicates the best result at each target-data fraction.
}
\label{tab:stage1_input_ablation}
\small
\setlength{\tabcolsep}{5.0pt}
\begin{tabular}{llcccc}
\toprule
Transfer & Stage-1 input & 10\% & 25\% & 50\% & 100\% \\
\midrule
$U\rightarrow U$
& Scratch
& 0.137${\pm}$0.049
& 0.152${\pm}$0.080
& 0.163${\pm}$0.120
& 0.192${\pm}$0.146 \\
\midrule
\multirow{3}{*}{$J\rightarrow U$}
& Full
& \textbf{0.454${\pm}$0.056}
& \textbf{0.434${\pm}$0.048}
& \textbf{0.426${\pm}$0.077}
& \textbf{0.413${\pm}$0.063} \\
& Coarse-only
& 0.324${\pm}$0.121
& 0.325${\pm}$0.085
& 0.357${\pm}$0.109
& 0.362${\pm}$0.065 \\
& Neural-only
& 0.310${\pm}$0.038
& 0.299${\pm}$0.060
& 0.315${\pm}$0.042
& 0.290${\pm}$0.040 \\
\midrule

\multirow{3}{*}{$W\rightarrow U$}
& Full
& 0.371${\pm}$0.072
& 0.387${\pm}$0.067
& \textbf{0.419${\pm}$0.076}
& \textbf{0.427${\pm}$0.043} \\
& Coarse-only
& \textbf{0.374${\pm}$0.066}
& \textbf{0.390${\pm}$0.063}
& 0.395${\pm}$0.081
& 0.399${\pm}$0.065 \\
& Neural-only
& 0.217${\pm}$0.111
& 0.223${\pm}$0.123
& 0.261${\pm}$0.109
& 0.202${\pm}$0.112 \\
\bottomrule
\end{tabular}
\end{table}

%% file: tables/dual-source.tex
\begin{table}[h]
\centering
\caption{
Dual-source cross-dataset transfer under different target-data fractions. Template denotes direct cross-dataset trajectory transfer without EEG. Results are trajectory PCC (mean$\pm$std); bold indicates the best result at each target-data fraction.
}
\label{tab:dual_source_transfer}
\small
\setlength{\tabcolsep}{4.5pt}
\begin{tabular}{llcccc}
\toprule
Dataset & Adaptation & 10\% & 25\% & 50\% & 100\% \\
\midrule

\multirow{4}{*}{Joystick-Track}
& \multicolumn{5}{l}{\textit{Template, Zero-shot PCC: -0.262 $\pm$ 0.113}} \\
& Scratch
& \textbf{0.730${\pm}$0.087}
& \textbf{0.702${\pm}$0.104}
& \textbf{0.733${\pm}$0.096}
& \textbf{0.742${\pm}$0.109} \\
& Full-FT
& 0.717${\pm}$0.101
& 0.674${\pm}$0.118
& 0.721${\pm}$0.097
& 0.729${\pm}$0.110 \\
& Stage-1-only
& -0.064${\pm}$0.140
& 0.035${\pm}$0.248
& 0.073${\pm}$0.241
& 0.245${\pm}$0.243 \\
\midrule

\multirow{4}{*}{WAY-EEG-GAL}
& \multicolumn{5}{l}{\textit{Template, Zero-shot PCC: 0.062 $\pm$ 0.180}} \\
& Scratch
& \textbf{0.757${\pm}$0.094}
& 0.738${\pm}$0.107
& \textbf{0.787${\pm}$0.083}
& 0.787${\pm}$0.112 \\
& Full-FT
& 0.717${\pm}$0.092
& 0.743${\pm}$0.113
& 0.780${\pm}$0.090
& 0.775${\pm}$0.097 \\
& Stage-1-only
& 0.729${\pm}$0.171
& \textbf{0.789${\pm}$0.102}
& 0.782${\pm}$0.135
& \textbf{0.816${\pm}$0.079} \\
\midrule

\multirow{4}{*}{U-Limb}
&  \multicolumn{5}{l}{\textit{Template, Zero-shot PCC: 0.175 $\pm$ 0.079}} \\
& Scratch
& 0.137${\pm}$0.049
& 0.152${\pm}$0.080
& 0.163${\pm}$0.120
& 0.192${\pm}$0.146 \\
& Full-FT
& 0.305${\pm}$0.139
& 0.300${\pm}$0.148
& 0.326${\pm}$0.133
& \textbf{0.363${\pm}$0.059} \\
& Stage-1-only
& \textbf{0.333${\pm}$0.062}
& \textbf{0.341${\pm}$0.080}
& \textbf{0.368${\pm}$0.071}
& 0.349${\pm}$0.062 \\
\bottomrule
\end{tabular}
\end{table}

%% file: tables/scaling.tex
\begin{table}[h]
\centering
\caption{Trajectory reconstruction performance under balanced trial scaling.
Values report the mean $\pm$ standard deviation of trajectory PCC over five common random seeds. The Macro column is the unweighted average across the three datasets.}
\label{tab:trial_scaling}
\small
\begin{tabular}{rcccc}
\toprule
\textbf{Training trials} &
\textbf{U-Limb} &
\textbf{WAY-EEG-GAL} &
\textbf{Joystick} &
\textbf{Macro} \\
\midrule
90  & 0.359 $\pm$ 0.055 & 0.657 $\pm$ 0.172          & 0.459 $\pm$ 0.160          & 0.492 $\pm$ 0.082 \\
180  & 0.306 $\pm$ 0.130 & 0.728 $\pm$ 0.051          & 0.629 $\pm$ 0.067          & 0.554 $\pm$ 0.069 \\
270  & 0.344 $\pm$ 0.074 & \textbf{0.747 $\pm$ 0.006} & 0.650 $\pm$ 0.037          & 0.580 $\pm$ 0.033 \\
360 & \textbf{0.380 $\pm$ 0.042} &
      0.731 $\pm$ 0.030 &
      \textbf{0.673 $\pm$ 0.043} &
      \textbf{0.595 $\pm$ 0.029} \\
\bottomrule
\end{tabular}
\end{table}

%% file: tables/objvt-cdata.tex
\begin{table*}[t]
\centering
\caption{Cross-dataset reconstruction under Full-FT. Each cell reports PCC ($\uparrow$) / RMSE ($\downarrow$), as mean $\pm$ std. Best and second-best results are bolded and underlined.}
\label{tab:cross_dataset_reconstruction}

\begingroup
\scriptsize
\setlength{\tabcolsep}{3pt}
\renewcommand{\arraystretch}{1.10}

\resizebox{\textwidth}{!}{%
\begin{tabular}{llccccc}
\toprule
Transfer & Data & AMRT & Vel-Int & Direct-Pos & Pos-Res & Pos-Vel \\
\midrule

\rowcolor{gray!8}
J$\rightarrow$U & 10\% &
\metricpair{\textbf{0.34$\pm$0.14}}{\textbf{93.06$\pm$15.22}} &
\metricpair{0.01$\pm$0.07}{101.74$\pm$11.67} &
\metricpair{0.07$\pm$0.10}{97.50$\pm$11.31} &
\metricpair{\underline{0.12$\pm$0.09}}{98.52$\pm$10.43} &
\metricpair{0.11$\pm$0.11}{\underline{94.41$\pm$12.05}} \\

\rowcolor{gray!8}
& 25\% &
\metricpair{\textbf{0.31$\pm$0.14}}{\textbf{91.92$\pm$16.41}} &
\metricpair{0.06$\pm$0.10}{99.36$\pm$11.98} &
\metricpair{0.06$\pm$0.05}{97.54$\pm$11.26} &
\metricpair{0.07$\pm$0.05}{98.59$\pm$10.34} &
\metricpair{\underline{0.09$\pm$0.10}}{\underline{95.65$\pm$10.56}} \\

\rowcolor{gray!8}
& 50\% &
\metricpair{\textbf{0.41$\pm$0.13}}{\textbf{87.05$\pm$15.37}} &
\metricpair{0.08$\pm$0.05}{100.42$\pm$11.71} &
\metricpair{0.05$\pm$0.07}{\underline{96.16$\pm$13.94}} &
\metricpair{0.06$\pm$0.03}{98.67$\pm$10.77} &
\metricpair{\underline{0.12$\pm$0.10}}{96.96$\pm$10.82} \\

\rowcolor{gray!8}
& 100\% &
\metricpair{\textbf{0.30$\pm$0.05}}{\textbf{89.99$\pm$11.78}} &
\metricpair{0.06$\pm$0.07}{101.65$\pm$17.16} &
\metricpair{0.15$\pm$0.11}{\underline{93.23$\pm$14.69}} &
\metricpair{\underline{0.19$\pm$0.16}}{93.49$\pm$14.48} &
\metricpair{0.19$\pm$0.09}{96.20$\pm$10.42} \\

\midrule

W$\rightarrow$U & 10\% &
\metricpair{\textbf{0.26$\pm$0.14}}{\textbf{92.80$\pm$9.84}} &
\metricpair{0.02$\pm$0.11}{102.39$\pm$12.00} &
\metricpair{0.08$\pm$0.06}{97.61$\pm$11.27} &
\metricpair{\underline{0.10$\pm$0.05}}{98.69$\pm$10.33} &
\metricpair{0.03$\pm$0.08}{\underline{96.55$\pm$13.16}} \\

& 25\% &
\metricpair{\textbf{0.26$\pm$0.18}}{\textbf{91.24$\pm$10.79}} &
\metricpair{0.05$\pm$0.08}{99.49$\pm$9.17} &
\metricpair{0.07$\pm$0.07}{\underline{97.24$\pm$10.94}} &
\metricpair{0.08$\pm$0.07}{98.68$\pm$10.41} &
\metricpair{\underline{0.09$\pm$0.12}}{97.57$\pm$11.33} \\

& 50\% &
\metricpair{\textbf{0.38$\pm$0.04}}{\textbf{88.99$\pm$9.25}} &
\metricpair{0.08$\pm$0.07}{100.59$\pm$11.22} &
\metricpair{\underline{0.11$\pm$0.11}}{\underline{94.89$\pm$11.93}} &
\metricpair{0.07$\pm$0.05}{98.70$\pm$10.37} &
\metricpair{0.08$\pm$0.08}{96.17$\pm$10.39} \\

& 100\% &
\metricpair{\textbf{0.30$\pm$0.15}}{\textbf{90.08$\pm$9.48}} &
\metricpair{0.03$\pm$0.06}{100.86$\pm$13.28} &
\metricpair{\underline{0.18$\pm$0.15}}{\underline{90.96$\pm$10.58}} &
\metricpair{0.17$\pm$0.18}{94.25$\pm$8.73} &
\metricpair{0.17$\pm$0.13}{92.28$\pm$10.78} \\

\midrule

\rowcolor{gray!8}
J$\rightarrow$W & 10\% &
\metricpair{\underline{0.74$\pm$0.08}}{\textbf{7.28$\pm$0.40}} &
\metricpair{\textbf{0.74$\pm$0.05}}{28.21$\pm$10.31} &
\metricpair{0.63$\pm$0.06}{\underline{8.11$\pm$0.93}} &
\metricpair{0.73$\pm$0.10}{8.41$\pm$1.00} &
\metricpair{0.63$\pm$0.06}{8.13$\pm$0.76} \\

\rowcolor{gray!8}
& 25\% &
\metricpair{\textbf{0.77$\pm$0.10}}{\textbf{6.89$\pm$0.49}} &
\metricpair{\underline{0.74$\pm$0.06}}{26.66$\pm$7.82} &
\metricpair{0.64$\pm$0.06}{\underline{7.85$\pm$0.61}} &
\metricpair{0.74$\pm$0.11}{8.09$\pm$0.53} &
\metricpair{0.64$\pm$0.06}{7.91$\pm$0.73} \\

\rowcolor{gray!8}
& 50\% &
\metricpair{\textbf{0.78$\pm$0.10}}{\textbf{7.00$\pm$0.92}} &
\metricpair{0.72$\pm$0.05}{26.80$\pm$7.52} &
\metricpair{0.64$\pm$0.07}{7.77$\pm$0.84} &
\metricpair{\underline{0.74$\pm$0.12}}{8.02$\pm$0.95} &
\metricpair{0.64$\pm$0.08}{\underline{7.76$\pm$0.73}} \\

\rowcolor{gray!8}
& 100\% &
\metricpair{\textbf{0.78$\pm$0.10}}{\textbf{6.94$\pm$0.92}} &
\metricpair{0.72$\pm$0.04}{21.77$\pm$4.47} &
\metricpair{0.63$\pm$0.08}{\underline{7.80$\pm$0.86}} &
\metricpair{\underline{0.74$\pm$0.10}}{8.06$\pm$0.96} &
\metricpair{0.64$\pm$0.07}{7.81$\pm$0.75} \\

\midrule

U$\rightarrow$W & 10\% &
\metricpair{0.69$\pm$0.08}{\textbf{8.02$\pm$0.69}} &
\metricpair{\textbf{0.73$\pm$0.06}}{34.63$\pm$10.33} &
\metricpair{0.63$\pm$0.06}{\underline{8.10$\pm$0.73}} &
\metricpair{\underline{0.69$\pm$0.08}}{8.31$\pm$0.83} &
\metricpair{0.55$\pm$0.24}{8.18$\pm$1.07} \\

& 25\% &
\metricpair{\textbf{0.75$\pm$0.11}}{\textbf{7.21$\pm$0.78}} &
\metricpair{0.72$\pm$0.06}{33.59$\pm$9.87} &
\metricpair{0.64$\pm$0.05}{\underline{8.18$\pm$0.81}} &
\metricpair{\underline{0.75$\pm$0.11}}{8.49$\pm$0.79} &
\metricpair{0.58$\pm$0.20}{8.31$\pm$1.14} \\

& 50\% &
\metricpair{\textbf{0.78$\pm$0.09}}{\textbf{6.74$\pm$0.42}} &
\metricpair{0.72$\pm$0.05}{25.99$\pm$10.23} &
\metricpair{0.64$\pm$0.06}{\underline{7.93$\pm$0.83}} &
\metricpair{\underline{0.73$\pm$0.12}}{8.22$\pm$0.89} &
\metricpair{0.63$\pm$0.05}{8.03$\pm$0.66} \\

& 100\% &
\metricpair{\textbf{0.79$\pm$0.08}}{\textbf{6.89$\pm$0.57}} &
\metricpair{0.69$\pm$0.09}{21.07$\pm$4.77} &
\metricpair{0.63$\pm$0.05}{\underline{7.94$\pm$0.72}} &
\metricpair{\underline{0.74$\pm$0.10}}{8.18$\pm$0.78} &
\metricpair{0.57$\pm$0.20}{8.17$\pm$1.04} \\

\midrule

\rowcolor{gray!8}
U$\rightarrow$J & 10\% &
\metricpair{\textbf{0.70$\pm$0.12}}{\textbf{9703.04$\pm$1253.21}} &
\metricpair{\underline{0.70$\pm$0.11}}{10500.09$\pm$1244.62} &
\metricpair{0.63$\pm$0.08}{\underline{10177.78$\pm$1077.18}} &
\metricpair{0.66$\pm$0.08}{10279.47$\pm$1651.33} &
\metricpair{0.65$\pm$0.08}{10291.93$\pm$1492.34} \\

\rowcolor{gray!8}
& 25\% &
\metricpair{\textbf{0.69$\pm$0.10}}{\textbf{9410.92$\pm$1413.53}} &
\metricpair{\underline{0.68$\pm$0.08}}{\underline{9644.25$\pm$909.67}} &
\metricpair{0.62$\pm$0.10}{10244.69$\pm$1848.47} &
\metricpair{0.65$\pm$0.09}{10869.92$\pm$2445.26} &
\metricpair{0.63$\pm$0.10}{10163.27$\pm$1919.24} \\

\rowcolor{gray!8}
& 50\% &
\metricpair{\textbf{0.73$\pm$0.09}}{\textbf{9069.12$\pm$1390.21}} &
\metricpair{\underline{0.72$\pm$0.09}}{\underline{10442.21$\pm$2346.37}} &
\metricpair{0.57$\pm$0.11}{10764.08$\pm$1862.79} &
\metricpair{0.61$\pm$0.12}{11946.61$\pm$1835.37} &
\metricpair{0.51$\pm$0.23}{10940.64$\pm$1585.02} \\

\rowcolor{gray!8}
& 100\% &
\metricpair{\textbf{0.73$\pm$0.13}}{\textbf{8890.17$\pm$1734.64}} &
\metricpair{\underline{0.72$\pm$0.13}}{\underline{9431.47$\pm$1907.78}} &
\metricpair{0.50$\pm$0.20}{11435.91$\pm$2202.48} &
\metricpair{0.60$\pm$0.16}{13251.42$\pm$1155.62} &
\metricpair{0.61$\pm$0.07}{10814.59$\pm$1647.86} \\

\midrule

W$\rightarrow$J & 10\% &
\metricpair{\textbf{0.73$\pm$0.08}}{\textbf{9644.84$\pm$1184.78}} &
\metricpair{\underline{0.72$\pm$0.07}}{10600.47$\pm$1979.67} &
\metricpair{0.65$\pm$0.09}{{9855.84$\pm$1272.79}} &
\metricpair{0.66$\pm$0.07}{10339.65$\pm$1594.82} &
\metricpair{0.65$\pm$0.10}{\underline{9648.75$\pm$1235.56}} \\

& 25\% &
\metricpair{\textbf{0.70$\pm$0.12}}{\textbf{9472.83$\pm$1706.06}} &
\metricpair{\underline{0.69$\pm$0.10}}{10033.63$\pm$1576.23} &
\metricpair{0.65$\pm$0.09}{\underline{9707.29$\pm$1741.14}} &
\metricpair{0.66$\pm$0.09}{10250.25$\pm$2337.83} &
\metricpair{0.63$\pm$0.09}{9775.11$\pm$1331.71} \\

& 50\% &
\metricpair{\textbf{0.72$\pm$0.10}}{\textbf{9120.93$\pm$1549.30}} &
\metricpair{\underline{0.71$\pm$0.11}}{\underline{9450.10$\pm$1726.37}} &
\metricpair{0.65$\pm$0.08}{9707.08$\pm$1397.09} &
\metricpair{0.66$\pm$0.06}{10019.41$\pm$1944.99} &
\metricpair{0.64$\pm$0.08}{9927.88$\pm$1853.48} \\

& 100\% &
\metricpair{\underline{0.73$\pm$0.11}}{\underline{8818.90$\pm$1549.65}} &
\metricpair{\textbf{0.73$\pm$0.10}}{\textbf{8771.44$\pm$1221.29}} &
\metricpair{0.64$\pm$0.09}{10026.05$\pm$1826.73} &
\metricpair{0.63$\pm$0.10}{10504.23$\pm$2593.62} &
\metricpair{0.63$\pm$0.08}{10022.70$\pm$1566.19} \\

\bottomrule
\end{tabular}%
}
\endgroup
\end{table*}

%% file: tables/model-ablation.tex
\begin{table}[h]
\centering
\caption{
Model ablations under the \texttt{ctri-mix} protocol.
Results are trajectory PCC (mean$\pm$std); bold indicates the best result.
}
\label{tab:model_ablation}
\small
\setlength{\tabcolsep}{5pt}
\begin{tabular}{lccc}
\toprule
Model & U-Limb & WAY-EEG-GAL & Joystick-Track \\
\midrule
\model{}  & \textbf{0.530$\pm$0.029}  &  \textbf{0.837$\pm$0.045}  & \textbf{0.787$\pm$0.064} \\
\model{}-Stage0  & 0.471$\pm$0.020  &  0.785$\pm$0.038  & 0.770$\pm$0.076 \\
w/o BiLSTM          & 0.364$\pm$0.038 & 0.831$\pm$0.018 & 0.725$\pm$0.024 \\
w/o Multi-Horizon   & 0.330$\pm$0.037 & 0.768$\pm$0.095 & 0.720$\pm$0.031 \\
w/o Gate  &0.502$\pm$0.022 &0.802$\pm$0.041 &0.727$\pm$0.036\\
w/o Gate-Increment  &0.497$\pm$0.021 &0.793$\pm$0.042 &0.705$\pm$0.022 \\
Pooled Decoder      & 0.385$\pm$0.022 & 0.832$\pm$0.027 & 0.740$\pm$0.023 \\
\bottomrule
\end{tabular}
\end{table}

%% file: tables/grasp.tex
\begin{table}[h]
\centering
\caption{
Grasp-state decoding performance on WAY-EEG-GAL under different evaluation protocols. We report accuracy and F1 score as mean$\pm$std.}
\label{tab:grasp_decoding}
\small
\setlength{\tabcolsep}{8pt}
\begin{tabular}{lcc}
\toprule
Protocol & Accuracy & F1 \\
\midrule
\texttt{ctri}     & 0.905$\pm$0.030 & 0.848$\pm$0.066 \\
\texttt{ctri-mix} & 0.868$\pm$0.040 & 0.794$\pm$0.056 \\
\texttt{csub}     & 0.883$\pm$0.031 & 0.844$\pm$0.047 \\
\bottomrule
\end{tabular}
\end{table}